\documentclass[aps,pre,twocolumn,showpacs,superscriptaddress,groupedaddress,reprint]{revtex4-1}

\usepackage{subfigure}
\usepackage{wrapfig}
\usepackage{textcomp}
\usepackage{graphicx}
\usepackage{amssymb}
\usepackage{amsmath}
\usepackage{bm}
\usepackage{amsmath, amsthm, amssymb}
\usepackage{hyperref}
\usepackage{color}
\usepackage{rotating}

\usepackage{amsfonts}
\usepackage{dcolumn}
\usepackage{float}
\usepackage{bm}

\newcommand {\ws} {\ensuremath{a}}

\newcommand {\rada} {\ensuremath{a_1}}
\newcommand{\radb} {\ensuremath{a_2}}
\newcommand{\radj} {\ensuremath{a_j}}
\newcommand{\radi} {\ensuremath{a_i}}
\newcommand{\abar}{\ensuremath{\left <A \right >}}

\newcommand{\qbar}{\ensuremath{\left <Q \right >}}

\newcommand{\udens}{\ensuremath{\,\text{g/cm}^3}}

\newcommand{\rmix}{\ensuremath{\rho_\text{mix}}}
\newcommand{\ISO}{\ensuremath{\text{iso-}n_e}}

\begin{document}
\setcounter{table}
{0}\title{Checking the Salpeter enhancement of nuclear reactions in asymmetric mixtures}

\author{ Jean Cl\'erouin, Philippe Arnault,  and Nicolas Desbiens}
\affiliation{CEA, DAM, DIF, 91297 Arpajon, France}
\author{Alexander J. White, Lee A. Collins, Joel D. Kress, and Christopher Ticknor }
\affiliation{Theoretical Division, Los Alamos National Laboratory, Los Alamos, New Mexico 87545, USA}

 \date{\today}

\begin{abstract}
We investigate the plasma enhancement of nuclear reactions in the intermediate coupling regime using orbital free molecular dynamics  (OFMD) simulations. Mixtures of H-Cu and H-Ag serve as prototypes of  simultaneous weak and strong couplings  due to the charge asymmetry. Of particular importance is the partial ionization of Cu and Ag and the free electron polarization captured by OFMD simulations. By comparing a series of OFMD simulations at various concentrations and constant pressure to multi-component hyper netted chain (MCHNC) calculations of effective binary ionic mixtures (BIM), we set a general procedure for computing enhancement factors. The MCHNC procedure allows extension to very low concentrations (5\% or less) and to very high temperatures (few keV) unreachable by the simulations. Enhancement factors for nuclear reactions rates extracted from the MCHNC approach are compared with the Salpeter theory in the weak and strong coupling regimes, and a new interpolation is proposed.
\end{abstract}

\maketitle
\section{ Introduction}
Since the seminal work of Salpeter \cite{SALP54,Salpeter1969}, it is recognized that fusion reaction rates in stars are enhanced by the plasma environment in every stage of their evolution from main sequence stars to red giants, and eventually for the cooling of white dwarfs and the explosions of supernovae. To first order, the two-body Coulomb interaction between the reacting nuclei is not modified and the plasma effect is mainly a constant lowering of the many-body Coulomb barrier, enhancing quantum  tunneling. This enhancement depends strongly on the physical conditions of the plasma, whether weakly or strongly coupled as determined by the ratio between the Coulomb interaction and thermal energy. The strength of this coupling must be considered when the stars evolve across coupling regimes. The enhancement depends also on the composition of the plasma, which varies from hydrogen-rich to degenerate carbon-oxygen cores. In addition, young stars and  giant planets mix hydrogen with helium and traces of heavier elements while white dwarf stars mix carbon with other high atomic number $Z$ elements. Supernovae are subjected to  violent shock-driven explosions creating  heavy elements that blend with primordial hydrogen. 
 These systems fall in the regime of matter under extreme conditions and warm dense matter (WDM). Mixtures also play a significant role in inertial confinement fusion (ICF). For example, mixing of the plastic ablator into the fuel has been used to partially explain lower than expected yields in ICF experiments \cite{SMAL13,EDWA13,ROBE13,RIND14}.  
  \begin{figure}[t]
\begin{center}
\includegraphics[width=6cm]{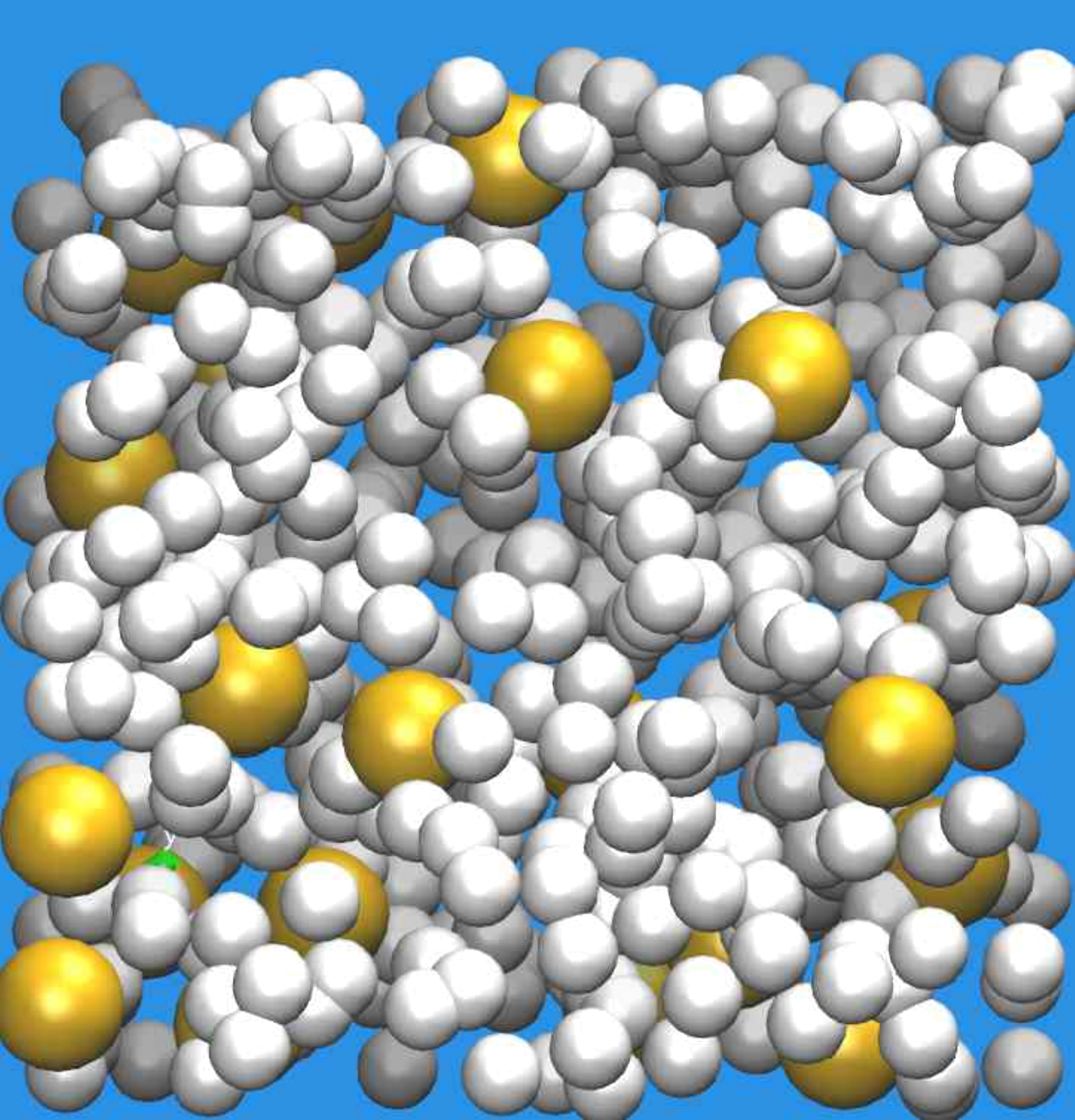}
\caption{\label{fig1} Snapshot extracted from an orbital free molecular dynamics simulation of a mixture of hydrogen (grey) with 5\% in number of copper (yellow)   at 100\,eV and 6.6\udens. Copper ions that are ionized 10 times more than hydrogen are drawn with a bigger radius.}
\end{center}
\end{figure}

Salpeter \cite{SALP54,Salpeter1969} proposed two formulations to account for the plasma enhancement, which are valid in the limiting cases of weak and strong coupling. Both models apply equally to plasma mixtures. Salpeter called the model for weak coupling ``weak screening", which causes enhancement of fusion reactions in, for instance, main sequence stars. He called the model for strong coupling "strong screening", which increases the fusion reaction rates by several orders of magnitude, mainly in white dwarf stars.  

The term ``screening" should refer to the shielding of ion--ion interactions by the electron polarization. At weak coupling (``weak screening" model), both the electrons and the ions act to enhance fusion reaction rates through the Debye screening lengths. In contrast, for white dwarf stars for instance, the electrons are highly degenerate and form a uniform rigid background (jellium). In this case ("strong screening" model), the electron screening does not contribute to the enhancement of fusion reactions. Many theoretical attempts have been devoted to model the transition between  weak and strong coupling \cite{DeWitt1973,Graboske1973,Mitler1977,Ogata1991,Rosenfeld1992,Ichimaru1993,Ogata1996,Kitamur2000,Yakovlev2006,Chugunov2009,Potekhin2012}. In the intermediate coupling regime typical of WDM, many challenging questions appear: How does the electron polarization evolve across coupling regimes \cite{Potekhin2012}? How does one account for partial ionization of spectators \cite{Taguchi1992}? Is there a peculiar regime associated with asymmetric mixtures \cite{WHIT15} where the highest charged ions behave as a strongly coupled plasma and the lowest charged ions as a weakly coupled plasma? 

As tools to investigate these issues, molecular dynamics simulations, a snapshot of which is shown in Fig.\,\ref{fig1}, and the Monte Carlo calculations are particularly demanding since resolution of the small $r$ behavior of the pair distribution function $g(r)$ is statistically difficult. The one component plasma (OCP) has been studied extensively providing many analytical formula across coupling regimes. To a lesser extent,  binary ionic mixtures (BIM) have also been studied with applications focused on the carbon-oxygen cores of white dwarf stars. More asymmetric mixtures, with charge ratios up to 44 have recently been examined \cite{WHIT15}. To our knowledge, no simulations exist which account for the polarization of the free electrons around the ions, nor any simulations performed where the ionization of the atoms can adapt to the physical conditions of the plasma. The present state of the art leaves many unanswered questions.

In this paper, we present orbital free molecular dynamics (OFMD) simulations of asymmetric mixtures: hydrogen--copper  (H-Cu) and hydrogen-silver (H-Ag). We first compared H-Cu and H-Ag at 100~eV to investigate the effect of a difference in the atomic number of the impurity. We choose these highly asymmetric mixtures in order to adequately challenge our analytical models. We extended our study with H-Ag mixtures at 100, 400, and 1600~eV to explore lower plasma couplings. We show that, while the structure of  the single component systems can be reproduced by the effective one component plasma (eOCP) \cite{CLER16}, the structure of the mixture is well described by the effective binary ionic mixture (eBIM) counterpart. Since a simple formulation for BIM structure is not available, we use a multi-component hyper-netted chain (MCHNC) estimation \cite{HANS06,ROGE80}.
Based on to the satisfactory agreement between simulations and MCHNC calculations, we used the MCHNC method  to predict enhancement factors for nuclear reactions between hydrogen nuclei. This procedure allows us to overcome the limitations of numerical simulations and to consider very low heavy element concentrations, \emph{e.g.} 1\%, and to investigate a wider range of temperatures up to a few keV. Estimates of enhancement factors versus heavy element  concentration at constant temperature  and pressure  are compared with the Salpeter formulations for weak  and strong couplings.

For the sake of clarity, we define some general notation. We consider binary mixtures of $N_1$ atoms of a light element of atomic number  $Z_1$ and mass $A_1$, and $N_2$ atoms of heavier atoms ($Z_2$,  $A_2$) with $N=N_1+N_2$. We will refer to the number concentration of the heavy element by $x=N_2/N$. The total volume $V$ sets the total density, $\rmix$,  and the average distance between atoms, the Wigner-Seitz radius, given in atomic units by  $\ws=\left [ 3/(4 \pi n) \right ]^{1/3}/a_B$, where $a_\text{B}$ is the Bohr radius.  $n={\cal N}\rmix/\abar$ is the total number density, $\abar=(1-x)A_1+xA_2$ the average atomic mass and ${\cal N}$ the Avogadro number.  The Wigner-Seitz radius $\ws$ and the temperature $T$, when expressed in atomic units, define the charge independent coupling parameter of the mixture by  $\Gamma_0=(\ws T_{au})^{-1}$, with $T_{au}=T_{eV}/27.2$. 

After a brief review of the Salpeter theory (Sec.\,\ref{salptheory}), some details are given on MCHNC calculations in Sec.\,\ref{mchnccalculations}, and on OFMD simulations in Sec.\,\ref{ofmdsimulations}. Both approaches are compared with respect to the ionic structure in Sec.\,\ref{struct}. The determination of the enhancement factor is presented in Sec.\,\ref{enhancement} and compared with Salpeter's models in Sec.\,\ref{concentration}.
\section{Salpeter's theory}
\label{salptheory}
Here, we review Salpeter's formulations at weak and strong couplings \cite{SALP54,Salpeter1969}, and we propose a generalization for partially ionized plasma mixtures.

The cross section $\sigma(E)$ of nonresonant fusion reactions between nuclei of charge $Z_i$ and $Z_j$ at low energy $E$ can be expressed as
\begin{equation}
\sigma(E) = \dfrac{S(E)}{E}\, \exp\left[-2 \pi\, \eta(E)\right],
\end{equation}
where the astrophysical factor $S(E)$ is a slowly varying function of $E$ and $\eta(E)$ arises from the quantum tunneling through the Coulomb barrier. The reactivity $\left <\sigma v \right >$ is obtained by averaging over the Maxwellian distribution for the relative motion, the product of the cross section $\sigma$ and a particle flux. The reaction rate per unit of volume is: 
\begin{equation}
R = \dfrac{1}{1+\delta_{ij}}\,n_i n_j \int_0^\infty\, dE\, S(E) \exp\left[ -2 \pi \eta(E) - \dfrac{E}{k T}\right],
\end{equation}
where the Kronecker symbol $\delta_{ij}$ excludes double counting of the collisions in reactions between identical nuclei of density $n_i = n_j$. In many cases, the integrand of $R$ presents a strong peak, named after Gamow, formed by the product of the increasing cross section $\sigma$ and the decreasing Maxwellian. 

Salpeter considered the first-order correction to the Coulomb potential due to the plasma environment, which is a constant lowering $(-H_0)$. Since the reaction occurs then in a potential shifted downward by $H_0$, the calculation proceeds as if the relative energy $E$ were shifted upward by $H_0$ in an unshifted Coulomb potential. The reaction rate with account of plasma effect reads
\begin{equation}
R_S = \dfrac{1}{1+\delta_{ij}}\,n_i n_j\, I_S,
\end{equation}
with
\begin{equation}
I_S = \int_0^\infty\, dE\, S(E+H_0) \exp\left[ -2 \pi \eta(E+H_0) - \dfrac{E}{k T}\right].
\end{equation}
Performing a change of integration variable leads to
\begin{equation}
I_S = \int_{H_0}^\infty\, dE\, S(E) \exp\left[ -2 \pi \eta(E) - \dfrac{E-H_0}{k T}\right].
\end{equation}
The lower integration limit $H_0$ can be shifted to $0$ if the Gamow peak is narrow enough and does not contribute in that range. The rate is then enhanced by a factor $f_S$ given by
\begin{equation}
f_S = \dfrac{R_S}{R} = \exp\left(\dfrac{H_0}{k T}\right).
\end{equation}

At very high coupling, some further corrections are needed that correspond to the next order in $(r^2)$ of the small-$r$ expansion of the screening potential $H_{ij}(r)$, the limit of which is $H_0$ at vanishing $r$ \cite{Ogata1991,Chugunov2009,Kravchuk2014}. 

The second step evaluates $H_0$ in a mean field approximation. This approach has been developed further by DeWitt \textit{et al.} \cite{DeWitt1973,Graboske1973}, Jancovici \cite{Jancovici1977}, and Ogata  \textit{et al.} \cite{Ogata1991} who showed that the small $r$ behavior of the pair distribution function (PDF) $g_{ij}(r)$ gives access to the corresponding expansion of the screening potential $H_{ij}(r)$
\begin{equation}
\label{hncgdr}
g_{ij}(r) = \exp\left[- \dfrac{V_{ij}(r) - H_{ij}(r)}{k T}\right],
\end{equation}
where $V_{ij}(r)$ is the \emph{direct} pair interaction potential between the reacting nuclei. Under the assumption of complete ionization, which will be discussed below, this potential involves the atomic numbers $Z_i$ and $Z_j$
\begin{equation}
V_{ij}(r) = \dfrac{Z_i Z_j\, e^2}{r}.
\end{equation}
The PDF is the ratio of the number density of species $i$ to the ideal gas density at a distance $r$ away from a particle of species $j$. In particle simulations, it is obtained averaging histograms accumulated on several snapshots according to
\begin{equation}
g_{ij}(r) = \dfrac{V}{N_i (N_j-\delta_{ij})}\,\left<\sum_p^{N_i}\,\mathop{{\sum}'}_{q}^{N_j}\,\delta(\mathbf{r}-\mathbf{r}_{pq}^{(ij)})\right>,
\end{equation}
where the primed sum excludes $q = p$ if $i=j$.
In classical statistical physics, the PDF is given for a single species plasma by
\begin{equation}
g(r) = \dfrac{N (N-1)}{n^2\,\mathcal{Z}}\,\int\,d\mathbf{r}_3 \dots d\mathbf{r}_N \, \exp\left(-\dfrac{U}{k T}\right),
\end{equation}
where $U$ is the sum of all pair interactions, and $\mathcal{Z}$ is the configurational integral associated with the excess free energy $A$ of the system
\begin{equation}
\mathcal{Z} = \int\,d\mathbf{r}_1 \dots d\mathbf{r}_N \, \exp\left(-\dfrac{U}{k T}\right) = \exp\left(-\dfrac{A}{k T}\right).
\end{equation}

The preceding authors \cite{DeWitt1973,Graboske1973,Jancovici1977,Ogata1991} elaborated on an argument originally due to Widom \cite{Widom1963} and showed in particular that $H_0$ is related to the difference between the excess free energy before and after the fusion reaction, \textit{i.e.} when the nuclei of charge $Z_i$ and $Z_j$ have formed a compound nucleus of charge $(Z_i + Z_j)$. At the very high densities of white dwarf stars and neutron star outer envelopes, quantum effects between the reacting and spectator nuclei introduce additional corrections \cite{Ogat1997,Pollock2004,Militzer2005,Chugunov2007}.  

\subsection{Salpeter's formulation at weak coupling}

For weak coupling, the potential of mean force between the nuclei is often assumed to be of the Debye-H\"uckel type
\begin{equation}
V_{ij}(r) - H_{ij}(r) = \dfrac{Z_i Z_j\, e^2}{r}\,\exp\left(- \dfrac{r}{\lambda}\right),
\end{equation}
where $\lambda$ is the total Debye screening length
\begin{equation}
\dfrac{1}{\lambda^2} = \sum_i \dfrac{1}{\lambda_i^2} + \dfrac{1}{\lambda_e^2}.
\end{equation}
Each species $i$ contributes with its own Debye length
\begin{equation}
\dfrac{1}{\lambda_i^2} = \dfrac{4 \pi\, n_i Z_i^2\,e^2}{k T},
\end{equation}
and the screening length of the electrons can differ from the ion expression at non-negligible degeneracy. In any case, the value of $H_0$ is given either from the free energy difference or simply from the difference between the Coulomb and  Debye-H\"uckel potentials
\begin{eqnarray}
H_0 &=& \lim_{r \to 0} \left[\dfrac{Z_i Z_j\, e^2}{r} - \dfrac{Z_i Z_j\, e^2}{r}\,\exp\left(- \dfrac{r}{\lambda}\right)\right], \\ \nonumber
    &=& \dfrac{Z_i Z_j\, e^2}{\lambda}.
\end{eqnarray}
The Debye approximation is inadequate at short distance for the electrons since their density can not be obtained in linear response close to a nucleus. Brown and  Sawyer \cite{BROW97} recently proposed an original and more physically sound derivation of the weak coupling Salpeter result.

\subsection{Salpeter's formulation at strong coupling}

For strong coupling, Salpeter approximated the free energy difference by assigning the value of $H_0$ by a difference of interaction energies. He also assumed that the electrons are completely degenerate and form a uniform background.
In such a situation, a plasma mixture can be viewed as a collection of neutral spheres of radii $a_j$ (ion sphere model), filled with a constant density of electrons neutralizing the nucleus of charge $Z_j$. The radius $\radi$ of each sphere depends on each nuclear charge $Z_j$ and ensures that the volume per electron stays constant throughout the system
\begin{equation}
\label{isoNeRule}
\dfrac{4 \pi\, \radj^3}{3\,Z_j} = \dfrac{V}{\sum_i N_i Z_i} = \dfrac{4 \pi\, a^3}{3\,\left<Z\right>},
\end{equation}
where $\left<Z\right> = \sum_i x_i Z_i$ is the mean charge of the nuclei ($x_i = N_i/N$).
The interaction energy of each "pseudo-atom" is
\begin{equation}
W(Z_j)=-{\frac {9}{10}}{\frac{Z_j^2\, e^2}{\radj}}.
\end{equation}
Therefore, $H_0$ is given by the energy difference between the composite pseudo-atom of nucleus charge ($Z_i+Z_j$) and the pseudo-atoms of nucleus charges $Z_i$ and $Z_j$.
\begin{equation}
H_0= W(Z_i+Z_j)-W(Z_i)- W(Z_j).
\end{equation}

\subsection{Partial ionization formulation}
\label{salpeterpartial}
In many cases at intermediate coupling, the plasma is partially ionized. Whereas the fusion reaction ultimately occurs between the nuclei of two completely stripped atoms, the spectator ions of the plasma retain their bound electrons and their cloud of polarized free electrons. This problem is less acute for hydrogen at high temperature where $Q_1 \sim Z_1 = 1$ but could have some importance for C-C reactions. Besides, high-$Z$ impurities, like iron in the sun, rarely reach complete ionization. 

We generalize Salpeter's formulations empirically using the partial ionization $Q_i$ of species $i$ in place of $Z_i$. The evolution of the bound electrons and the polarization clouds around the reacting nuclei before and after the reaction is unresolved \textit{faute de mieux}. In the OFMD simulations, the close encounters are never close enough to probe this stage. However, the simulations probe as realistically as possible the effect of the spectator atoms.

To complete this generalization, the ionization $Q_i$ of each species $i$ needs to be modeled. We propose to use the average atom approximation. For each species $i$, we require that the electron density $n_e$ at the boundary of the confinement sphere of radius $a_i$ be the same and be shared by all species. This kind of modeling is called "iso-$n_e$" in the plasma community. If one considers that the electron density in excess of $n_e$ and the nucleus represents an effective ion of charge $Q_i(a_i,T)$, the neutrality of the average atom implies
\begin{equation}
Q_i(\radi,T) = \dfrac{4}{3}\,\pi\, \radi^3\,n_e.
\end{equation}
The different radii $\radi$ and ionizations $Q_i$ are then obtained solving Eq.\,\eqref{isoNeRule} iteratively with $Q_i(\radi,T)$ in place of $Z_i$, leading for the case of light--heavy mixtures (H-Cu or H-Ag) to
\begin{equation}
\label{iso}
{\frac {\rada}{Q_1^{1/3}}} = {\frac {\radb}{Q_2^{1/3}}} = {\frac {\ws}{\qbar^{1/3}}},
\end{equation}
where $\qbar=(1-x)Q_1+xQ_2 $ is the average charge. Starting from $\rada=\radb=\ws$, the solution is quickly reached, using More's fit \cite{MORE83,MORE85} of the  Thomas-Fermi average ionization  to determine $Q_1$ an $Q_2$ at each iteration. The effective ionizations $Q_i(\radi,T)$ include the polarization cloud of free electrons around the nucleus in addition to the bound electrons. This accounts implicitly for electron screening of the ion-ion interactions. Consequently, the explicit electron contribution to the enhancement factor at weak coupling must be discarded to avoid double-counting their screening effect.

When we analyze OFMD results using the MCHNC method, we make the additional assumption that the charges $Q_i(\radi,T)$ are point-like ions in an electron jellium. For pure element, this defines an effective OCP concept that has been shown to give a comprehensive picture of the short-range properties, such as the PDF, the equation of state and the transport coefficients \cite{CLER13b,ARNA13,CLER15,CLER16,CLER16b}. We shall see that this concept stays valid for plasma mixtures, defining an effective BIM representative of the real system as simulated with OFMD.

Defining an average over the charges by $\left<Q^n\right> = \sum_i x_i Q_i^n$, the generalized weak coupling Salpeter enhancement factor of the fusion reaction between nuclei of species $i$ and $j$ reads
\begin{equation}
h_W=\frac {H_0}{kT} = Q_i Q_j\, \Gamma_0\, \sqrt{3 \Gamma_0 \left<Q^2\right>}.
\label{salpW}
\end{equation}
The strong screening Salpeter's formulation is 
\begin{equation}
h_S=\frac {H_0}{kT} = \dfrac{9}{10}\,\Gamma_0 \left<Q\right>^{1/3} \left[(Q_i+Q_j)^{5/3}-Q_i^{5/3}-Q_j^{5/3}\right].
\label{salpS}
\end{equation}

For reactions between hydrogen atoms this leads to the following formula for strong and weak coupling regimes
\begin{eqnarray}
h_W&=& Q_1^2\, \Gamma_0\, \sqrt{3 \Gamma_0 \left<Q^2\right>}  \\
h_S&=& 1.057\,\Gamma_0 \left<Q\right>^{1/3} Q_1^{5/3}.
\end{eqnarray}

\section{Multi-component HNC}
\label{mchnccalculations}
 The HNC approach for a single species plasma is based on the general Ornstein-Zernike (OZ) relation
\begin{equation}
h(r)=c(r)+n \int c\left (  |\mathbf{r}-\mathbf{r}'| \right )h(\mathbf{r}') d\mathbf{r}',
\label{oz1}
\end{equation}
in which $h(r)=g(r)-1$ and $c(r)$ is the direct correlation function. The OZ relation introduces the direct correlation between two particles separated by $r$ (first term), but also accounts for the correlation due to all of the other pairs (second term). To determine $g(r)$ from a two-body potential $V(r)$ we use a closure relation
\begin{equation}
c(r)=-\beta V(r)+h(r)-\log\left[h(r)+1\right]+B(r),
\label{oz2}
\end{equation}
where $\beta=1/kT$.  It is useful to write Eq.~(\ref{oz2}) as
 \begin{eqnarray}
\beta H(r)&=&n \int c\left (  |\mathbf{r}-\mathbf{r}'| \right )h(\mathbf{r}') d\mathbf{r}'+B(r)	\\ \nonumber
		&=&\beta V(r)+\log(g(r)),
\label{oz3}
\end{eqnarray}
which emphasizes that the screening function $H(r)$, introduced in Eq.\,\eqref{hncgdr}, is an essential ingredient of both the theory of plasma enhancement of the nuclear reactions and the HNC integral equations.  $B(r)$ is the bridge function representing three-body and higher order correlations. In the original HNC scheme, $B(r)$ is neglected.  

 The generalization to system of $M$ different species leads to as many OZ relations and closures as pair interactions between species \cite{HANS06,ROGE80}
 \begin{eqnarray}
h_{ij}(r)&=&c_{ij}(r)+n \sum_{k =1}^M x_k \int c_{ik}\left (  |\mathbf{r}-\mathbf{r}'| \right )h_{kj}(\mathbf{r}') d\mathbf{r}' \nonumber  \\
c_{ij}(r)&=&-\beta V_{ij}(r)+h_{ij}(r)-\log\left[h_{ij}(r)+1\right]+B_{ij}(r), \nonumber \\
\label{oz1m}
\end{eqnarray}

The inputs of the  MCHNC calculations are the charges, concentration, density, and temperature. 
The charges $Q_1$ and $Q_2$ are obtained through the $\ISO$  prescription given by Eq.\,\eqref{iso} in Sec.\,\ref{salpeterpartial}.  
Table\,\ref{tab1} collects the input parameters for the MCHNC calculations in columns 4-6, for each concentration $x$  and  temperature.  Note the relative constancy of the ionizations $Q_1$ and $Q_2$ with respect to $x$. The different data associated with a given concentration $x$ of heavy atoms are constrained to be at the same pressure as confirmed by the OFMD pressure given in column 3. This constraint is almost equivalent to a requirement of constant electron density, since the total pressure is dominated by the electron pressure, and the electron pressure is mainly driven by the electron density. Therefore, our iso-$n_e$ procedure gives almost the same ionization whatever the concentration $x$. The only varying parameter is the coupling $\Gamma_0$  which depends  on the density at constant temperature. 

\begin{table}[!t]
\caption{\label{tab1}Parameters for OFMD simulations (2 first columns) and MCHNC calculations (a, $Q_1$, $Q_2$ from Eq. \ref{iso})  and resulting screening potential $H_0$ at vanishing $r$  for a H-Cu mixture at 100~eV  and H-Ag mixtures at 100, 400 and 1600\,eV, for various concentrations $x$. $\rho_\text{mix}$ is the density in \udens. Column 3 gives the  OFMD pressure in units of the corresponding pure hydrogen pressure at 2\udens.  Screening values in bold correspond to the OCP enhancement for pure hydrogen.$Gamma_0$ is the charge independent coupling parameter.}
\begin{center}
\begin{tabular}{|c|c|c||c|c|c|c|c|}
\hline  
\multicolumn{8}{|c|}{H-Cu 100 eV}\\
\hline
	$x$\%& $\rmix$	&	$P/P_H$	&	$\ws$	 &$Q_1$	& $Q_2$ &$10^2  \Gamma_0$	&$10^2 H_0/kT$	\\
\hline  
	0	&2.00	&	1.		&	1.10		&	0.919	&	-		&	24.6	&	{\bf 13.3}		\\
	5	&6.56	&	0.970	&	1.19		&	0.916	&	9.91		&	22.9	&	20.4	\\
	10	&9.60	&	0.968	&	1.26		&	0.914	&	9.93		&	21.5	&	23.6	\\
	25	&14.7	&	0.952	&	1.45		&	0.911	&	9.97		&	18.8	&	27.7	\\
	50	&18.3	&	0.939	&	1.68		&	0.909	&	10.01	&	16.2	&	30.1	\\
	75	&20.0	&	1.034	&	1.86		&	0.908	&	10.03	&	14.6	&	31.2	\\
	100	&21.0	&	0.928	&	2.00		&	-		&	10.05	&	13.5	&	-		\\
\hline 	\hline
\multicolumn{8}{|c|}{H-Ag 100 eV}\\
\hline 
	$x$\%& $\rmix$	&	$P/P_H$	&	$\ws$	 &$Q_1$	& $Q_2$ &$10^2  \Gamma_0$	&$10^2 H_0/kT$	\\
\hline 
	0	&2.00	&	1.		&	1.10		&	0.919	&	-		&	24.6	&	{\bf 13.3}	\\
	5	&9.60	&	0.989	&	1.21		&	0.915	&	11.93	&	22.5	&	21.9	\\
	10	&14.3	&	0.966	&	1.30		&	0.913	&	12.00	&	21.0	&	25.4	\\
	25	&21.5	&	0.949	&	1.51		&	0.910	&	12.11	&	18.0	&	29.7	\\
	50	&26.3	&	0.944	&	1.77		&	0.909	&	12.19	&	15.4	&	32.1	\\
	75	&28.5	&	0.941	&	1.97		&	0.908	&	12.23	&	13.8	&	33.1	\\
	100	&29.7	&	0.936	&	2.13		&	-		&	12.26	&	12.7	&	-	\\
\hline 	\hline
\multicolumn{8}{|c|}{H-Ag 400 eV}\\
\hline 
	$x$\%& $\rmix$	&	$P/P_H$	&	$\ws$	 &$Q_1$	& $Q_2$ &$10^2 \Gamma_0$	&$10^2 H_0/kT$	\\
\hline  
	0	&2.00	&	1.00		&	1.10		&	0.979	&	-		&	6.16	&	{\bf 2.3}	\\
	5	&7.95	&	1.01		&	1.29		&	0.977	&	24.47	&	5.28	&	7.1	\\
	10	&10.7	&	0.99		&	1.43		&	0.976	&	24.21	&	4.76	&	8.3	\\
	25	&14.0	&	0.97		&	1.74		&	0.975	&	23.94	&	3.90	&	9.3	\\
	50	&15.8	&	0.97		&	2.10		&	0.974	&	23.81	&	3.24	&	9.8	\\
	75	&16.5	&	0.97		&	2.36		&	0.974	&	23.76	&	2.88	&	9.9	\\
	100	&16.9	&	0.96		&	2.58		&	-		&	23.74	&	2.64	&	-	\\
\hline 	\hline
\multicolumn{8}{|c|}{H-Ag 1600 eV}\\
\hline \
	$x$\%& $\rmix$	&	$P/P_H$	&	$\ws$	 &$Q_1$	& $Q_2$ &$10^2 \Gamma_0$	&$10^2 H_0/kT$	\\
\hline  
	0	&2.00	&	1.		&	1.10		&	0.995	&	-		&	1.54	&	{\bf 0.3}	\\
	5	&6.37	&	0.997	&	1.38		&	0.994	&	40.27	&	1.23	&	1.7	\\
	10	&7.86	&	0.990	&	1.59		&	0.994	&	40.10	&	1.11	&	2.0	\\
	25	&9.36	&	0.987	&	2.00		&	0.994	&	39.96	&	0.85	&	2.2	\\
	50	&10.0	&	0.985	&	2.44		&	0.994	&	39.89	&	0.70	&	2.3	\\
	75	&10.3	&	0.984	&	2.76		&	0.994	&	39.87	&	0.62	&	2.3	\\
	100	&10.4	&	0.984	&	3.02		&	-		&	39.85	&	0.56	&	-	\\
	\hline	
\end{tabular}
\end{center}
\end{table}

\section{ OFMD simulations}
\label{ofmdsimulations}
Simulating mixtures like H-Cu or H-Ag at hundreds of eVs  and about 2-30\udens, where the heavy component cannot be considered as fully ionized, is not a simple task. In these conditions, quantum molecular dynamics simulations based on the Kohn-Sham representation are particularly challenging, due to the large number of orbitals required.  On the other hand, classical molecular dynamics simulations would be tractable, but need an {\it a priori} knowledge of ionization, and more generally of interactions between  species. To ensure the {\it ab initio} aspect of the simulations, we used OFMD simulations \cite{CLER92,LAMB07} that are particularly well adapted to this range of thermodynamic conditions and use no empirical parameters, \emph{e.g.} ionization. The method relies on the finite temperature Thomas-Fermi functional for electron kinetic energy which does not require orbitals, but only the density, in the spirit of the original Hohenberg-Kohn theorem. OFMD allows the study of plasmas of any atomic number with a constant computational cost in a wide range of temperature and density.

 \begin{figure}[t]
\includegraphics[width=7 cm]{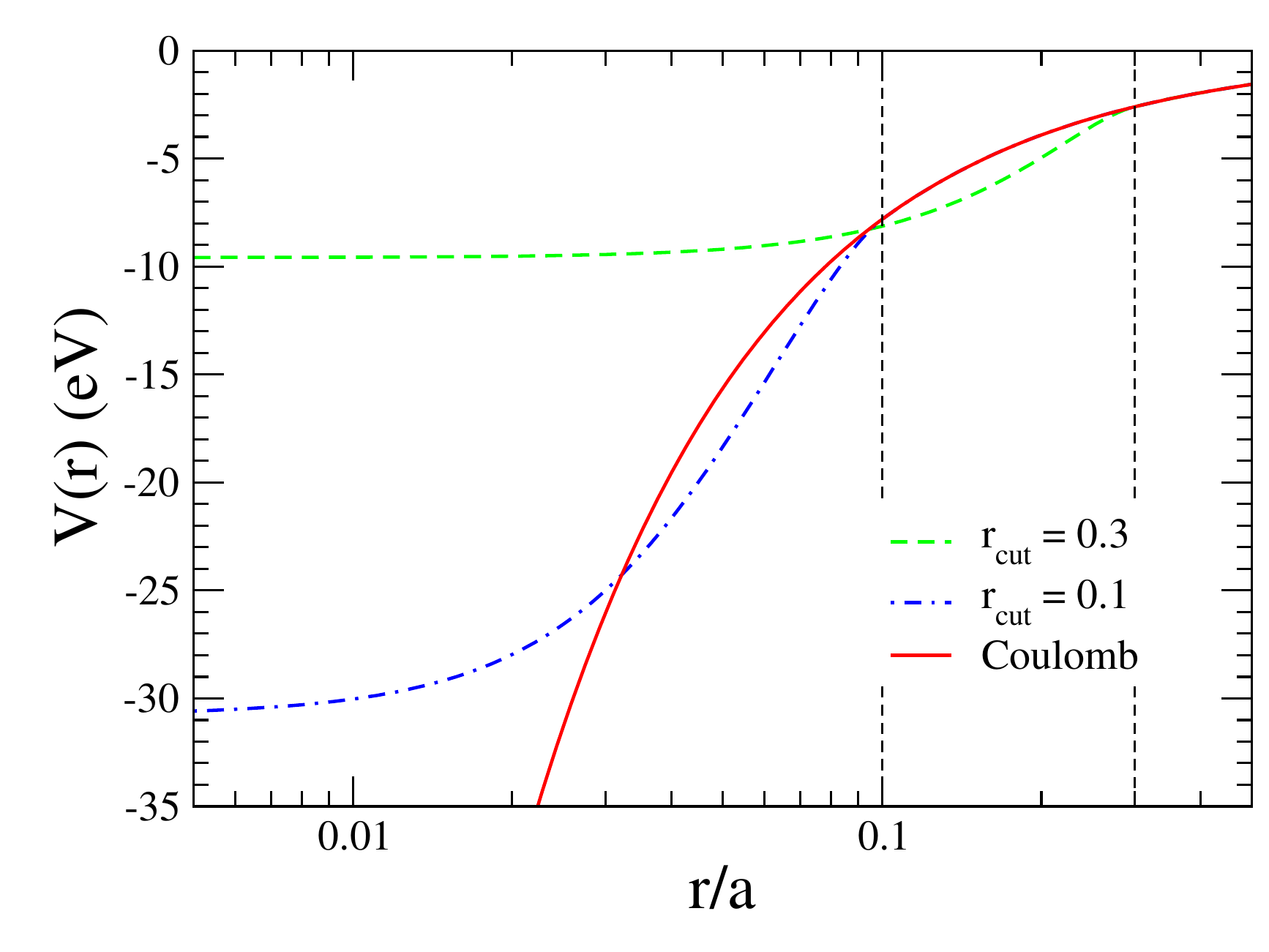}
\caption{\label{fig2}  (Color online) Comparison between the electron-nucleus bare Coulomb potential (red solid line) and the regularized potential with a cutoff radius $r_\text{cut}$  of $0.3\ws$ (green dashed line) or $0.1\ws$ (blue dot dashed  line).} 
\end{figure}

 We ran OFMD simulations with the Perrot finite-temperature exchange-correlation functional \cite{PERR79}  for H-Cu and H-Ag mixtures  at various concentrations and temperatures. H-Cu and H-Ag are compared at 100~eV and little effect due to  a difference in the atomic number of the impurity is found. Lower plasma couplings were explored with H-Ag mixtures at 100, 400, and 1600~eV.   We used a total number $N=432$ particles with a time step ranging from 2 to 0.5 atomic unit of time (a.u.) at 1600~eV. A time step of 10~a.u. would be sufficient to accurately  treat pure Ag at 1600~eV  due to the strong interactions between highly charged silver atoms. However, an accurate simulation of mixtures with hydrogen atoms requires smaller time-steps, by at least one order of magnitude, to account for   the weakly coupled regime of hydrogen.   To set the temperature,  simulations were performed  in the isokinetic ensemble, which has proven to be very efficient in equilibrium situations.   
 
  In the present context, it must be emphasized  that the electron-nucleus potential must be regularized at short distance, to avoid divergencies of the electron density in the Thomas-Fermi description, which leads to a Fourier expansion of the density over a prohibitive number of components \cite{LAMB07}. This norm-conserving pseudo-potential introduces a cutoff radius $r_\text{cut}$, which is usually taken as  $r_\text{cut}=0.3 \ws$. Below a distance of  $2r_\text{cut}$, the ion-ion  interactions are affected by this regularization.  We show in Fig.~\ref{fig2}, that this regularized potential is slightly lower than the coulomb one between $r_\text{cut}/3$ and $r_\text{cut}$ and goes to a constant below $r_\text{cut}/3$. For relative distances $r< 2r_\text{cut}$, overlaps happen, but between $2 r_\text{cut}/3$ and $2r_
  c$ the distribution functions $g(r)$ are not strongly affected by this regularization. In contrast,  screening functions $H(r)$   are very sensitive to the short range behavior, and data below $2r_\text{cut}$ must be taken with care. 
 
\begin{figure}[!t]
\includegraphics[width=9cm]{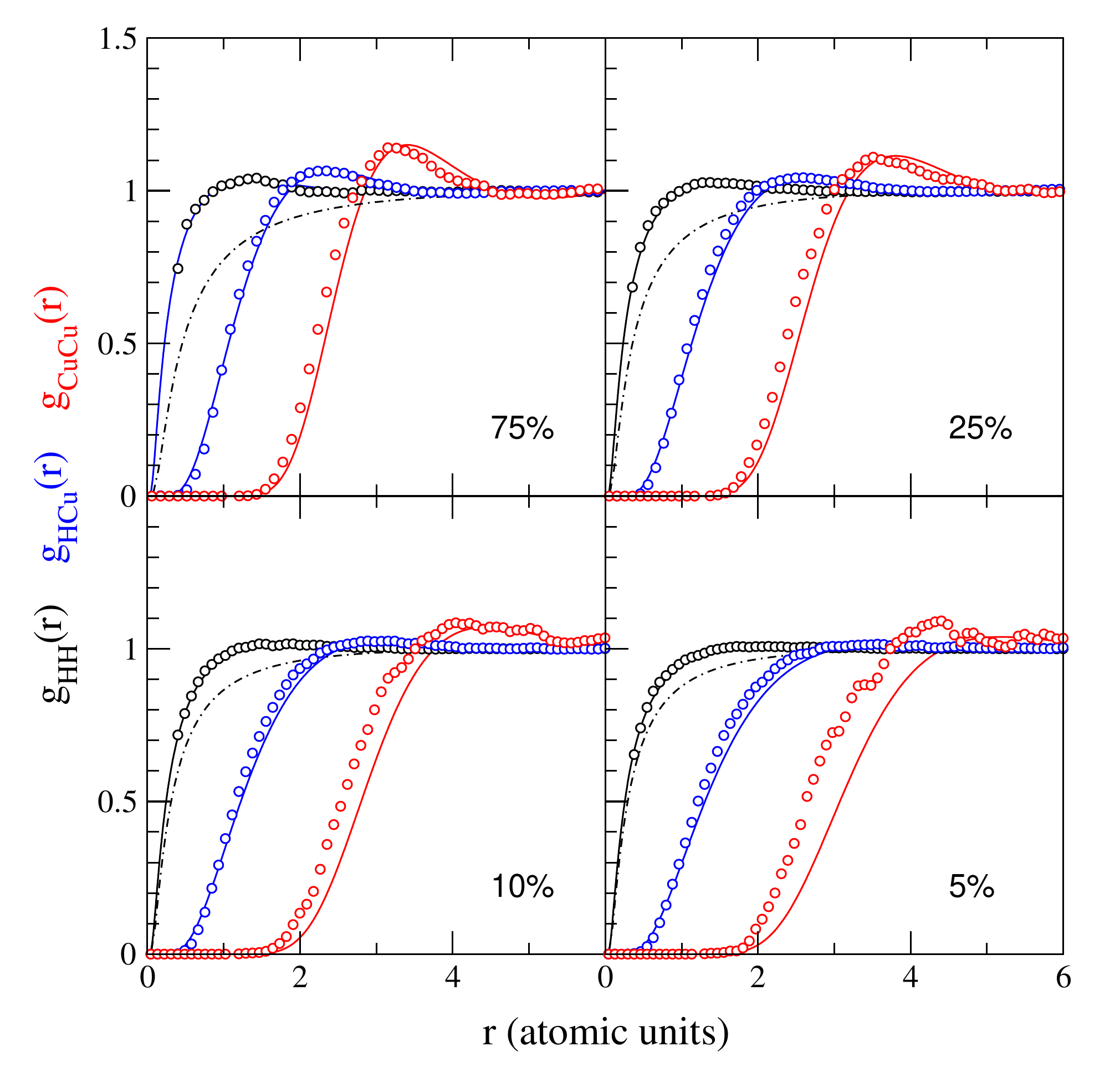}
\caption{\label{fig3} (Color online) Pair distribution functions of the H-Cu mixtures at 100~eV and concentration, $x$, from 5 to 75\%, see Table \ref{tab1}. From left to right, in black: hydrogen-hydrogen; blue: hydrogen-copper; red: copper-copper. Open circles: OFMD simulations, solid lines: MCHNC computations. Dot dashed line corresponds to a pure hydrogen OCP   at $\Gamma = Q_1^2\,\Gamma_0=0.2$ scaled to atomic units using $a$.} 
\end{figure}
\begin{figure}[!h]
\includegraphics[width=9cm]{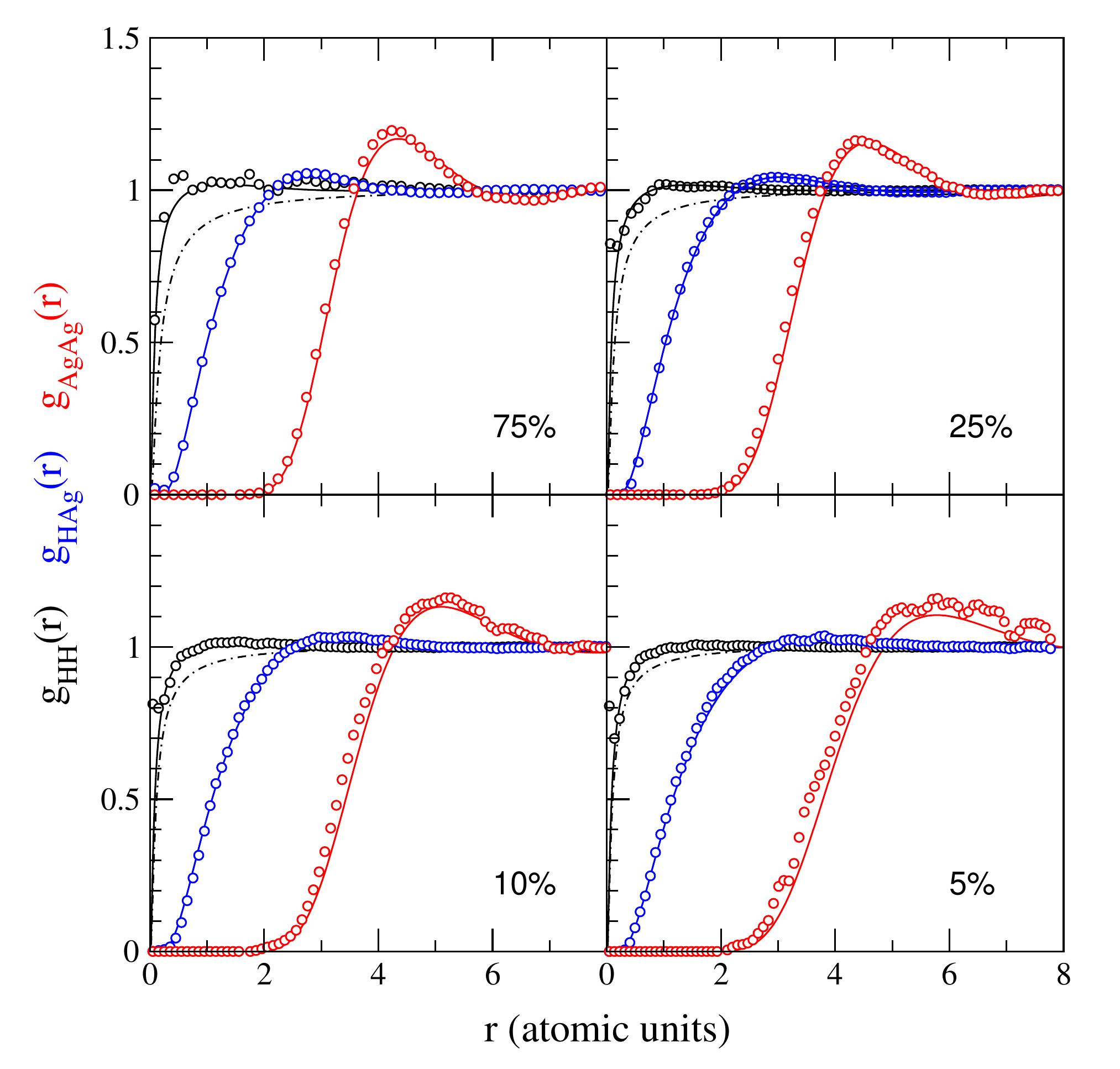}
\caption{\label{fig4} (Color online) Same as Fig.\,\ref{fig3} for the H-Ag mixtures  of Table\,\ref{tab1} at 400~eV. Dot dashed line corresponds to a pure hydrogen OCP   at  $\Gamma = Q_1^2\,\Gamma_0=0.06$ scaled to atomic units using $a$.} 
\end{figure}
A series of increasing concentrations have been performed following an isobaric rule, representative of the experimental conditions of mixing layers, e.g. in ICF targets. Concentrations of 0 (pure hydrogen), 5, 10, 25, 50, 75, and 100\% in the heavy element were considered at various temperatures, all at  the same pressure corresponding to the pressure  $P_H$ of pure hydrogen at 2\udens  (volume $V_1$).    To determine the mixture density \rmix\,at the same pressure, we search for the density of the pure heavy component that gives same pressure as the pure hydrogen, which defines a volume $V_2$, and use a simple composition rule
\begin{equation}
\rmix={\frac{(1-x)A_1+xA_2}{(1-x)V_1+x V_2}}.
\label{rhomix}
\end{equation}

Corresponding densities and  Wigner-Seitz radii  are given in Table\,\ref{tab1} for each concentration.  In this table,  the pressures of H-Cu and H-Ag mixtures at 100~eV  are very similar since in both cases we used the same reference pressure point (hydrogen at 2\udens). The slight variations of the pressures reached in OFMD simulations, can be traced back to the mixing rule of Eq.\,(\ref{rhomix}) and a simplified equation of state based on the OCP and the homogeneous electron gas \cite{TICK16}.
\begin{figure}[]
\includegraphics[width=7 cm]{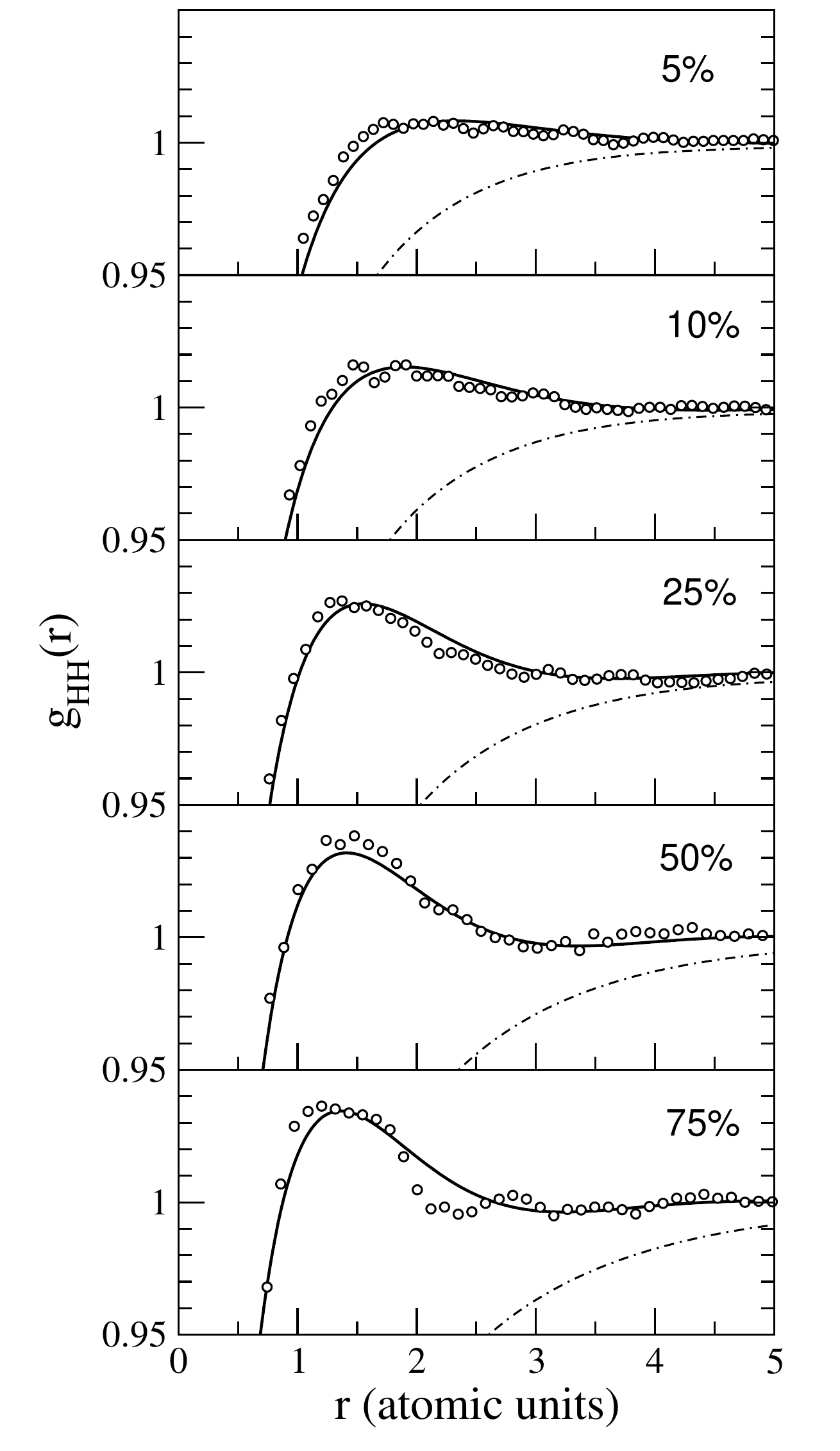}
\caption{\label{fig5} Enlargement of the peak of $g_{HH}$  as a function of Cu concentration (from top to bottom: 5, 10, 25, 50 and 75\%).  Circles are the OFMD results, the dot dashed lines are the effective OCP for pure hydrogen, at  $\Gamma = Q_1^2\,\Gamma_0=0.2$ scaled to atomic units using $\ws$, and black solid lines are the MCHNC computations with $Q_1$, $Q_2$ and $\ws$ given in Table\,\ref{tab1}. } 
\end{figure}

\section{Structure of the mixture}
\label{struct}
An overview of the structure of H-Cu and H-Ag mixtures is depicted in Fig.\,\ref{multi} for all concentrations and temperatures considered in Table\,\ref{tab1}. Fig.\,\ref{fig3} and Fig.\,\ref{fig4}  give details  for H-Cu at 100~eV and H-Ag at 400~eV.  The three PDFs, $g_\text{HH}$,  $g_\text{HZ}$, and $g_\text{ZZ}$ (Z=Cu or Ag) obtained by OFMD simulations  are shown for four selected concentrations $x$  (5, 10, 25 and 75\%).  Note that the 5\% concentration structure in Fig.~\ref{fig3} is the statistical analysis of Fig.~\ref{fig1}. The structure of the three pair distribution functions reflects the increasing strength of the interactions between ions. The Z-Z PDF exhibits a well defined peak which is also reproduced by an eOCP, with the heavy component being the dominant scatterer. On the contrary, it is not possible to find an  eOCP that matches the H-H structure due to the crowding effect on H atoms induced by the heavy elements. The difference in height between H-H PDF (black circles) and the OCP result (dot-dashed lines)  for pure hydrogen is a measure of this compression effect. To accurately describe the mixture, one needs to generalize the eOCP concept to an eBIM for mixtures. The  MCHNC model provides a way of computing such an eBIM, with effective ionizations $Q_1$, $Q_2$ given by some external model, such as the $\ISO$ prescription used   here. The values are given in Table~\ref{tab1}. With this choice of parameters ( columns 4-6 of Table\,\ref{tab1}) we get an excellent overall agreement between OFMD simulations and MCHNC calculations of the BIM as shown in Figs.\,\ref{fig3}-\ref{fig4} and in Fig.\,\ref{multi}. This confirms the pertinence of the $\ISO$ prescription within the BIM description.

An enlargement on the hydrogen structure in the presence of an increasing proportion of copper is shown in Fig.~\ref{fig5}. The over-compression clearly  appears when compared with the eOCP at  $\Gamma = Q_1^2\,\Gamma_0=0.2$, which corresponds to pure hydrogen at 2\udens\, and 100\,eV. This structure of overcorrelation is well reproduced by the MCHNC model even at large  scale, where statistical noise appears.  At low concentration  in the heavy material (5\%-10\%), we observe a mismatch  between the OFMD and MCHNC calculated $g_\text{ZZ}$ heavy correlation functions. This could be easily fixed by an empirical correction of the Wigner-Seitz radius, $\ws$, which scales MCHNC results. We identified this feature as originating partially  from the exchange and correlation corrections to the Thomas-Fermi functionals.  At  temperature around 100\,eV this mismatch amounts to an underestimation by about 10\% of the screening factors. Since this feature  vanishes with temperature (see Fig.\,\ref{fig4} and Fig.\,\ref{multi}), we did not correct for it in the temperature range of interest for fusion reactions ($T \gtrsim 1\,\text{keV}$). The agreement between OFMD simulations and MCHNC calculations motivates application of the MCHNC approach to compute nuclear enhancement factors. 

\begin{figure}[!h]
\includegraphics[width=6 cm]{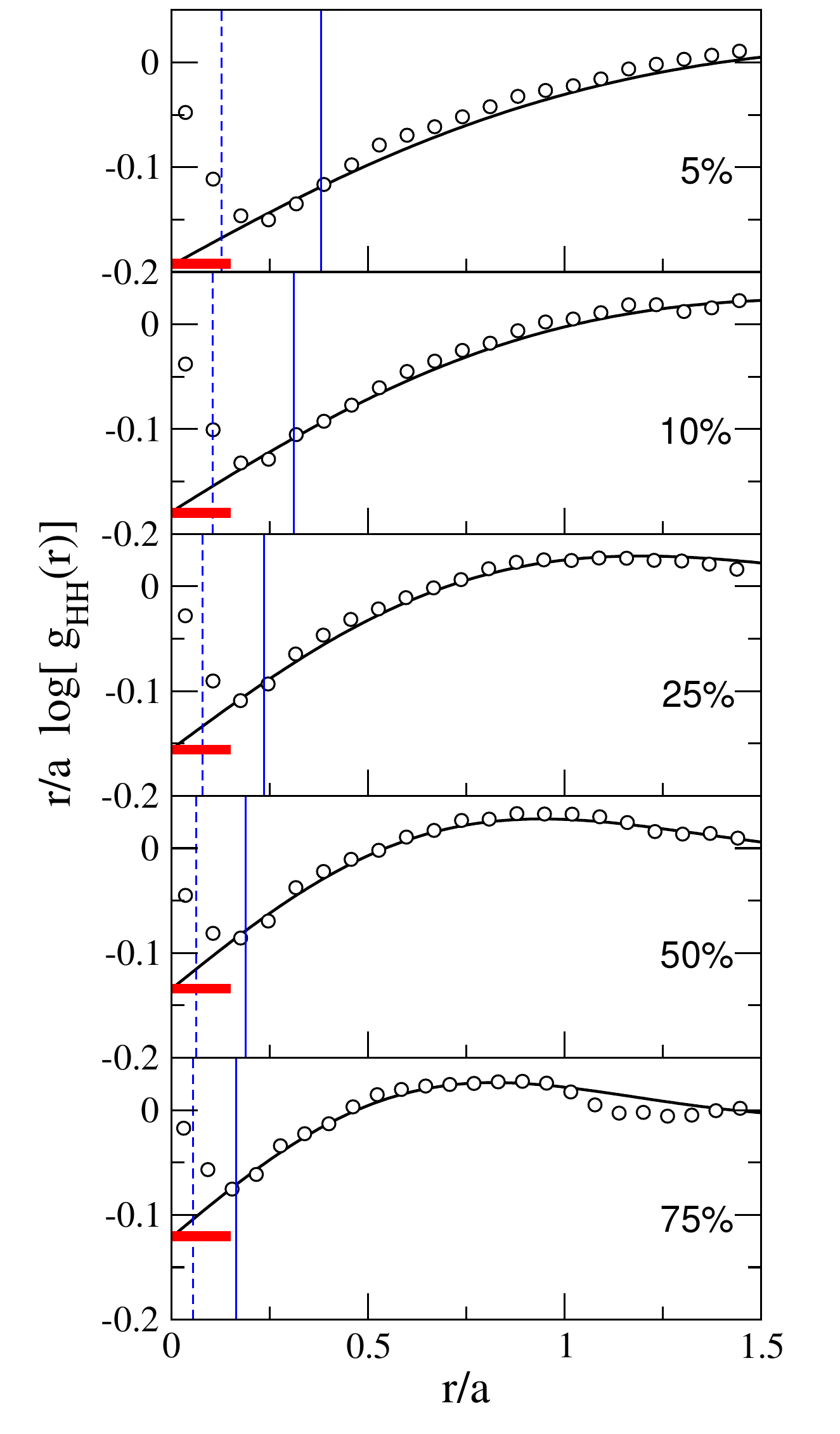}
\caption{\label{fig6}  Comparison of the function $r/a \log(g_\text{HH}(r))$ as obtained from OFMD simulations of the H-Cu mixture of Table\,\ref{tab1} at varied concentrations  (open circles) and MCHNC calculations (solid lines). This function intercepts the $y$ axis at ($-\Gamma=-Q_1^2\,\Gamma_0$)  (red segment on $y$ axis) with a slope equal to $H_0/kT$. The vertical blue lines correspond to the cutoff radius $2 r_\text{cut}$ (solid) and $2r_\text{cut}/3$ (dashed).  } 
\end{figure}

\section{Determination of the enhancement factor}
\label{enhancement}
The information on screening is expressed by the function
\begin{equation}
{\frac {H(r)}{kT}}  =  {\frac {\Gamma}{r}} +  \log[g(r)],
\label{defH}
\end{equation}
which is the difference between the bare coulomb potential ($\Gamma/r$) and the potential of mean force.
In section \ref{struct}, we saw that the MCHNC approach reproduces the overall structure given by  OFMD simulations. To check the compatibility of both approaches at vanishing distance,  we have plotted in Fig.\,\ref{fig6} the function $r\log(g(r))$, which intercepts the $y$ axis at $(-\Gamma=-Q_1^2 \Gamma_0)$ with a slope equal to $H_0/kT$. The agreement between both approaches is very good except at short distance ($r< 2 r_\text{cut}$), where OFMD points exhibits some scatter. As discussed in section \ref{ofmdsimulations}, the electron-nucleus potential must be regularized below $r_\text{cut}$. At  $r< 2r_\text{cut}$, overlaps happen,  which explains the scattered data in this range. The MCHNC calculation extrapolates the OFMD results at contact, which determines the enhancement factor $H_0/kT$. As a consequence, MCHNC allows one to investigate also situations too demanding for simulations, e.g. very low concentrations and high temperatures.

\section{Comparison with Salpeter's theory}
\label{concentration}
\begin{figure}[!t]
\includegraphics[width=7 cm]{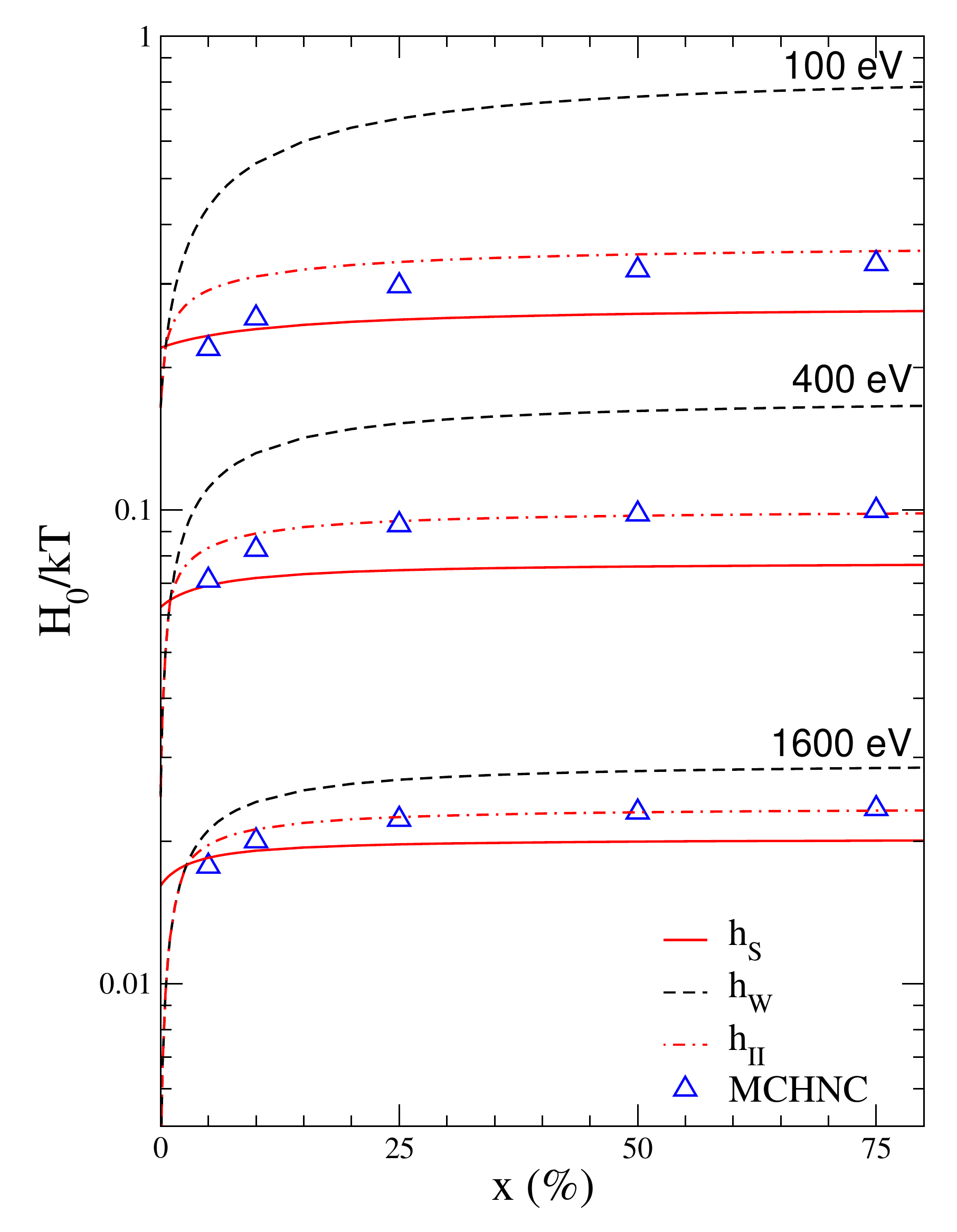}
\caption{\label{fig7}(Color online)  MCHNC values (blue triangles) of  the enhancement factor $H_0/kT$  for  the H-Ag  mixtures of Table\,\ref{tab1} at 100, 400 and 1600\,eV versus concentration, compared with the Salpeter formulations for strong coupling $h_S$  (red solid lines) and  weak coupling $h_W$ (black dashed lines). The red dot dashed lines correspond to the interpolation $h_{II}$ given by Eq.~(\ref{I})  with $\alpha=2$ and Eq.~(\ref{II}).}
\end{figure}

\begin{figure}[!t]
\includegraphics[width=7 cm]{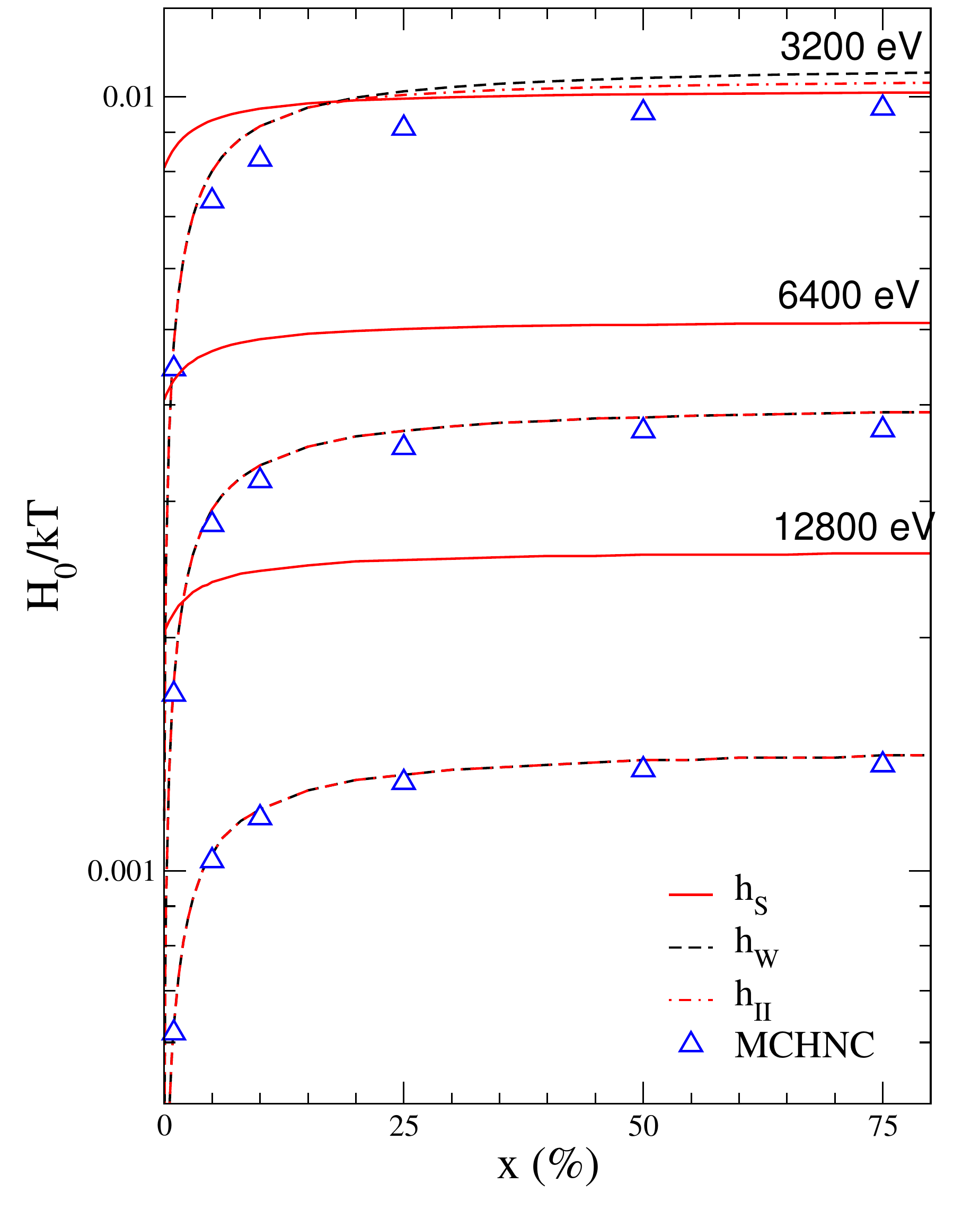}
\caption{\label{fig8} (Color online) Same as Fig.\,\ref{fig7} for the H-Ag mixtures of Table\,\ref{tab2} at temperatures of 3200, 6400 and 12800\,eV versus concentration $x$.} 
\end{figure}

All these comparisons substantiate a standard approach based on the $\ISO$ prescription to define effective charges feeding the MCHNC calculations of the corresponding BIMs. We compare in Figs.~\ref{fig7} and  ~\ref{fig8}   the estimates of the enhancement factor (blue triangles) with the Salpeter predictions. The value of $H_0/kT$ for a hydrogen-silver mixture is shown versus concentration $x$ for temperatures of 100, 400 and 1600\,eV in Figs.~\ref{fig7}  and 3200, 6400 and 12800\,eV in Fig.\,\ref{fig8}.  The input data  for Figs.~\ref{fig7} and \ref{fig8} are given in Table\,\ref{tab1} and \ref{tab2}, respectively.     For each temperature,  the black dashed line is the Salpeter  weak coupling formulation, $h_W$,  and the red solid line  Salpeter formulation for strong coupling, $h_S$. At 100\,eV,  $h_W$ strongly overestimates the results for all concentrations and is 2-3 times higher than the MCHNC results. In contrast, $h_S$ slightly underestimates them up to temperatures of 3200\,eV, where all predictions are equivalent. At higher temperatures, 6400 and 32800\,eV, the $h_S$ prediction becomes larger than $h_W$ and our results are in excellent agreement with the weak coupling formulation, $h_W$.

It is thus tempting to look for an interpolation formula. The standard interpolation is the harmonic average which selects the lowest value if they are sufficiently different,

\begin{equation}
h_I={\frac{h_Wh_S}{\left (h_W^{\alpha}+h_S^{\alpha}   \right )^{1/\alpha}}}.
\label{I}
\end{equation}
Salpeter \cite{Salpeter1969} proposed a quadratic interpolation ($\alpha=2$) but a quartic interpolation ($\alpha=4$) has been also proposed by Yakovlev \cite{Yakovlev2006}.  These formulations reproduce very well our data in the weakly-coupled regime but give a poorer estimation in the intermediate regime where both $h_W$ and $h_S$ are of the same order. 
To cover this regime, we correct the harmonic average at low temperature by a factor of $\sqrt{2}$, while retaining $h_W$ at high temperature

\begin{equation}
h_{II}=\text{Min} \left [ \sqrt{2}  h_I, h_W \right ].
\label{II}
\end{equation}
This interpolation adequately reproduces our data in the whole range of temperatures.

\begin{table}[!t]
\caption{\label{tab2}Parameters for  MCHNC calculations  (\ws, $Q_1$,  
$Q_2$)   and  screening potential $H_0$ at vanishing $r$  for  H-Ag mixtures at 3200, 6400 and 12800\,eV, for various concentrations $x$.  $\rho_\text{mix}$ is given in \udens. Screening values in bold correspond to the OCP enhancement for pure hydrogen. $Gamma_0$ is the charge independent coupling parameter.}
\begin{center}
\begin{tabular}{|c|c||c|c|c|c|c|}
\hline  
\multicolumn{7}{|c|}{H-Ag 3200 eV}\\
\hline 
	$x$\%& $\rmix$	&	$\ws$	 &$Q_1$	& $Q_2$ &$10^3 \Gamma_0$	&$10^3 H_0/kT$	\\
\hline 
	0	&2.00	&	1.10	&	0.997	&	-		&	7.7	&	{\bf 1.14}	\\
	0.1	&2.17	&	1.11	&	0.997	&	44.39	&	7.6	&	1.87	\\
	1	&3.40	&	1.18	&	0.997	&	44.28	&	7.2	&	4.45	\\
	5	&6.10	&	1.41	&	0.997	&	44.08	&	6.0	&	7.32	\\
	10	&7.42	&	1.64 &	0.997	&	44.02	&	5.3	&	8.19	\\
	25	&8.70	&	2.04	&	0.997	&	43.92	&	4.2	&	9.18	\\
	50	&9.28	&	2.50	&	0.997	&	43.89	&	3.4	&	9.53	\\
	75	&9.49	&	2.84	&	0.997	&	43.88	&	3.0	&	9.66	\\
	100	&9.61	&	3.11	&	-		&	43.87	&	2.7	&	-	\\
\hline 	\hline
\multicolumn{7}{|c|}{H-Ag 6400 eV}\\
\hline 
	$x$\%& $\rmix$	&	$\ws$	 &$Q_1$	& $Q_2$ &$10^3 \Gamma_0$	&$10^3 H_0/kT$	\\
\hline  
	0	&2.00		&	1.105	&	0.999	&	-	&	3.8	&	{\bf 0.40}	\\
	0.1	&2.17	&	1.113	&	0.999	&	45.88	&	3.8	&	0.70	\\
	1	&3.38	&	1.180	&	0.999	&	45.82	&	3.6	&	1.68	\\
	5	&6.02	&	1.413	&	0.999	&	45.73	&	3.0	&	2.80	\\
	10	&7.29	&	1.626	&	0.998	&	45.69	&	2.6	&	3.19	\\
	25	&8.52	&	2.058	&	0.998	&	45.65	&	2.1	&	3.52	\\
	50	&9.06	&	2.524	&	0.998	&	45.64	&	1.7	&	3.66	\\
	75	&9.27	&	2.862	&	0.998	&	45.63	&	1.5	&	3.71	\\
	100	&9.37	&	3.135	&	-		&	45.63	&	1.4	&	-	\\
\hline 	\hline
\multicolumn{7}{|c|}{H-Ag 12800 eV}\\
\hline \
	$x$\%& $\rmix$	&	$\ws$	 &$Q_1$	& $Q_2$ &$10^3 \Gamma_0$	&$10^3 H_0/kT$	\\
\hline  
	0	&2.00	&	1.11	&	0.999	&	-		&	1.9	&	{\bf 0.142}	\\
	0.1	&2.17	&	1.11	&	0.999	&	46.50	&	1.9	&	0.256	\\
	1	&3.38	&	1.18	&	0.999	&	46.47	&	1.8	&	0.628	\\
	5	&6.03	&	1.41	&	0.999	&	46.42	&	1.5	&	1.05	\\
	10	&7.30	&	1.63	&	0.999	&	46.40	&	1.3	&	1.17	\\
	25	&8.53	&	2.06	&	0.999	&	46.38	&	1.0	&	1.30	\\
	50	&9.08	&	2.52	&	0.999	&	46.38	&	0.8	&	1.35	\\
	75	&9.28	&	2.86	&	0.999	&	46.37	&	0.7	&	1.37	\\
	100	&9.39	&	3.13	&	-		&	46.37	&	0.7	&	-	\\
	\hline	
\end{tabular}
\end{center}
\end{table}

It is worth emphasizing that, at high temperature ($T> 3200\,eV$), screening factors are much smaller  than in the intermediate regime but are more sensitive to the presence of the heavy element. As a reference, we give in  Tables \ref{tab1} and \ref{tab2}, the screening factor for pure hydrogen (in bold). This value  is also fairly reproduced by Ichimaru's parametrization \cite{ICHI84}. When a slight amount of heavy element (0.1, 1 and 5\%)  is added,  the screening is increased by a factor of more than five for the highest temperatures, but only by a factor of two at lower temperatures.

\section{Conclusion}
We have addressed the plasma enhancement of nuclear reactions for asymmetric mixtures in the intermediate coupling regime of WDM, with partial ionization and free-electron polarization. The enhancement factor is obtained from the short range behavior of pair distribution function from orbital free molecular dynamics simulations. We considered mixtures of hydrogen at various concentrations of copper and silver at densities $2-30 \udens$, and temperatures $100-1600\,\text{eV}$. We find that an effective binary ionic mixture reproduces adequately the ionic structures provided that the effective ionizations are defined through an iso-electronic density prescription. This effective ionization captures the partial ionization and the free-electron polarization. The robustness of the eBIM concept is supported by our comparisons of two different chemical elements, Cu and Ag, mixed with H, and by the temperature dependency for different plasma coupling regimes. In particular, the eBIM description, computed with the MCHNC approach, reproduces the overcorrelation of the H subsystem, characteristic of the crowding effect due to the high $Z$ impurity. In practice, MCHNC calculations yield straightforwardly enhancement factors. Taking advantage of the excellent agreement between simulations and integral equations in the intermediate regime ($100 < T < 1600\, \text{eV}$), we pushed to the weak coupling regime ($T > 3200  \,\text{eV}$) and found  excellent agreement with the weak coupling Salpeter  prediction. In the intermediate regime, our results suggest a simple interpolation between weak and strong coupling, with generalized  Salpeter models for partial ionization. 
The effect of the addition of a high $Z$ element has been quantified in both regimes: The plasma enhancement is far higher in the weak coupling regime, and can scale $H_0/kT$ by almost one order of magnitude. 
The eBIM concept should find further applications in the description of WDM mixtures, such as equation of state and ionic transport coefficients \cite{CLER17,ARNA13b,TICK16,WHIT17}. More complex mixtures (ternary etc.) can be straightforwardly addressed by this effective multicomponent modeling.

%
%
\section{Acknowledgements}
PA, NB and JC wish to thank Serge Bouquet  and Ma\"elle Le Pennec for fruitful discussions. The authors gratefully acknowledge support from ASC, computing resource from CCC, and LANL which is operated by LANS, LLC for the NNSA of the U.S. DOE under Contract No. DE-AC52-06NA25396. This work has been performed under the NNSA/DAM collaborative agreement P184 on Basic Science. We specially thank Flavien Lambert for providing his OFMD code.


\begin{thebibliography}{44}
\expandafter\ifx\csname natexlab\endcsname\relax\def\natexlab#1{#1}\fi
\expandafter\ifx\csname bibnamefont\endcsname\relax
  \def\bibnamefont#1{#1}\fi
\expandafter\ifx\csname bibfnamefont\endcsname\relax
  \def\bibfnamefont#1{#1}\fi
\expandafter\ifx\csname citenamefont\endcsname\relax
  \def\citenamefont#1{#1}\fi
\expandafter\ifx\csname url\endcsname\relax
  \def\url#1{\texttt{#1}}\fi
\expandafter\ifx\csname urlprefix\endcsname\relax\def\urlprefix{URL }\fi
\providecommand{\bibinfo}[2]{#2}
\providecommand{\eprint}[2][]{\url{#2}}

\bibitem[{\citenamefont{Salpeter}(1954)}]{SALP54}
\bibinfo{author}{\bibfnamefont{E.~E.} \bibnamefont{Salpeter}},
  \bibinfo{journal}{Australian Journal of Physics}
  \textbf{\bibinfo{volume}{7}}, \bibinfo{pages}{373} (\bibinfo{year}{1954}).

\bibitem[{\citenamefont{Salpeter and Van~Horn}(1969)}]{Salpeter1969}
\bibinfo{author}{\bibfnamefont{E.~E.} \bibnamefont{Salpeter}} \bibnamefont{and}
  \bibinfo{author}{\bibfnamefont{H.~M.} \bibnamefont{Van~Horn}},
  \bibinfo{journal}{The Astrophysical Journal} \textbf{\bibinfo{volume}{155}},
  \bibinfo{pages}{183} (\bibinfo{year}{1969}).

\bibitem[{\citenamefont{Smalyuk et~al.}(2013)\citenamefont{Smalyuk, Atherton,
  Benedetti, Bionta, Bleuel, Bond, Bradley, Caggiano, Callahan, Casey
  et~al.}}]{SMAL13}
\bibinfo{author}{\bibfnamefont{V.~A.} \bibnamefont{Smalyuk}},
  \bibinfo{author}{\bibfnamefont{L.~J.} \bibnamefont{Atherton}},
  \bibinfo{author}{\bibfnamefont{L.~R.} \bibnamefont{Benedetti}},
  \bibinfo{author}{\bibfnamefont{R.}~\bibnamefont{Bionta}},
  \bibinfo{author}{\bibfnamefont{D.}~\bibnamefont{Bleuel}},
  \bibinfo{author}{\bibfnamefont{E.}~\bibnamefont{Bond}},
  \bibinfo{author}{\bibfnamefont{D.~K.} \bibnamefont{Bradley}},
  \bibinfo{author}{\bibfnamefont{J.}~\bibnamefont{Caggiano}},
  \bibinfo{author}{\bibfnamefont{D.~A.} \bibnamefont{Callahan}},
  \bibinfo{author}{\bibfnamefont{D.~T.} \bibnamefont{Casey}},
  \bibnamefont{et~al.}, \bibinfo{journal}{Phys. Rev. Lett.}
  \textbf{\bibinfo{volume}{111}}, \bibinfo{pages}{215001}
  (\bibinfo{year}{2013}),
  \urlprefix\url{https://link.aps.org/doi/10.1103/PhysRevLett.111.215001}.

\bibitem[{\citenamefont{Edwards et~al.}(2013)\citenamefont{Edwards, Patel,
  Lindl, Atherton, Glenzer, Haan, Kilkenny, Landen, Moses, Nikroo
  et~al.}}]{EDWA13}
\bibinfo{author}{\bibfnamefont{M.~J.} \bibnamefont{Edwards}},
  \bibinfo{author}{\bibfnamefont{P.~K.} \bibnamefont{Patel}},
  \bibinfo{author}{\bibfnamefont{J.~D.} \bibnamefont{Lindl}},
  \bibinfo{author}{\bibfnamefont{L.~J.} \bibnamefont{Atherton}},
  \bibinfo{author}{\bibfnamefont{S.~H.} \bibnamefont{Glenzer}},
  \bibinfo{author}{\bibfnamefont{S.~W.} \bibnamefont{Haan}},
  \bibinfo{author}{\bibfnamefont{J.~D.} \bibnamefont{Kilkenny}},
  \bibinfo{author}{\bibfnamefont{O.~L.} \bibnamefont{Landen}},
  \bibinfo{author}{\bibfnamefont{E.~I.} \bibnamefont{Moses}},
  \bibinfo{author}{\bibfnamefont{A.}~\bibnamefont{Nikroo}},
  \bibnamefont{et~al.}, \bibinfo{journal}{Physics of Plasmas}
  \textbf{\bibinfo{volume}{20}}, \bibinfo{pages}{070501}
  (\bibinfo{year}{2013}), \eprint{https://doi.org/10.1063/1.4816115},
  \urlprefix\url{https://doi.org/10.1063/1.4816115}.

\bibitem[{\citenamefont{Robey et~al.}(2013)\citenamefont{Robey, MacGowan,
  Landen, LaFortune, Widmayer, Celliers, Moody, Ross, Ralph, LePape
  et~al.}}]{ROBE13}
\bibinfo{author}{\bibfnamefont{H.~F.} \bibnamefont{Robey}},
  \bibinfo{author}{\bibfnamefont{B.~J.} \bibnamefont{MacGowan}},
  \bibinfo{author}{\bibfnamefont{O.~L.} \bibnamefont{Landen}},
  \bibinfo{author}{\bibfnamefont{K.~N.} \bibnamefont{LaFortune}},
  \bibinfo{author}{\bibfnamefont{C.}~\bibnamefont{Widmayer}},
  \bibinfo{author}{\bibfnamefont{P.~M.} \bibnamefont{Celliers}},
  \bibinfo{author}{\bibfnamefont{J.~D.} \bibnamefont{Moody}},
  \bibinfo{author}{\bibfnamefont{J.~S.} \bibnamefont{Ross}},
  \bibinfo{author}{\bibfnamefont{J.}~\bibnamefont{Ralph}},
  \bibinfo{author}{\bibfnamefont{S.}~\bibnamefont{LePape}},
  \bibnamefont{et~al.}, \bibinfo{journal}{Physics of Plasmas}
  \textbf{\bibinfo{volume}{20}}, \bibinfo{pages}{052707}
  (\bibinfo{year}{2013}), \eprint{https://doi.org/10.1063/1.4807331},
  \urlprefix\url{https://doi.org/10.1063/1.4807331}.

\bibitem[{\citenamefont{Rinderknecht et~al.}(2014)\citenamefont{Rinderknecht,
  Sio, Li, Zylstra, Rosenberg, Amendt, Delettrez, Bellei, Frenje, Gatu~Johnson
  et~al.}}]{RIND14}
\bibinfo{author}{\bibfnamefont{H.~G.} \bibnamefont{Rinderknecht}},
  \bibinfo{author}{\bibfnamefont{H.}~\bibnamefont{Sio}},
  \bibinfo{author}{\bibfnamefont{C.~K.} \bibnamefont{Li}},
  \bibinfo{author}{\bibfnamefont{A.~B.} \bibnamefont{Zylstra}},
  \bibinfo{author}{\bibfnamefont{M.~J.} \bibnamefont{Rosenberg}},
  \bibinfo{author}{\bibfnamefont{P.}~\bibnamefont{Amendt}},
  \bibinfo{author}{\bibfnamefont{J.}~\bibnamefont{Delettrez}},
  \bibinfo{author}{\bibfnamefont{C.}~\bibnamefont{Bellei}},
  \bibinfo{author}{\bibfnamefont{J.~A.} \bibnamefont{Frenje}},
  \bibinfo{author}{\bibfnamefont{M.}~\bibnamefont{Gatu~Johnson}},
  \bibnamefont{et~al.}, \bibinfo{journal}{Phys. Rev. Lett.}
  \textbf{\bibinfo{volume}{112}}, \bibinfo{pages}{135001}
  (\bibinfo{year}{2014}).

\bibitem[{\citenamefont{DeWitt et~al.}(1973)\citenamefont{DeWitt, Graboske, and
  Cooper}}]{DeWitt1973}
\bibinfo{author}{\bibfnamefont{H.~E.} \bibnamefont{DeWitt}},
  \bibinfo{author}{\bibfnamefont{H.~C.} \bibnamefont{Graboske}},
  \bibnamefont{and} \bibinfo{author}{\bibfnamefont{M.}~\bibnamefont{Cooper}},
  \bibinfo{journal}{The Astrophysical Journal} \textbf{\bibinfo{volume}{181}},
  \bibinfo{pages}{439} (\bibinfo{year}{1973}).

\bibitem[{\citenamefont{Graboske et~al.}(1973)\citenamefont{Graboske, DeWitt,
  Grossman, and Cooper}}]{Graboske1973}
\bibinfo{author}{\bibfnamefont{H.~C.} \bibnamefont{Graboske}},
  \bibinfo{author}{\bibfnamefont{H.~E.} \bibnamefont{DeWitt}},
  \bibinfo{author}{\bibfnamefont{A.~S.} \bibnamefont{Grossman}},
  \bibnamefont{and} \bibinfo{author}{\bibfnamefont{M.~S.}
  \bibnamefont{Cooper}}, \bibinfo{journal}{The Astrophysical Journal}
  \textbf{\bibinfo{volume}{181}}, \bibinfo{pages}{457} (\bibinfo{year}{1973}).

\bibitem[{\citenamefont{Mitler}(1977)}]{Mitler1977}
\bibinfo{author}{\bibfnamefont{H.~E.} \bibnamefont{Mitler}},
  \bibinfo{journal}{The Astrophysical Journal} \textbf{\bibinfo{volume}{212}},
  \bibinfo{pages}{513} (\bibinfo{year}{1977}).

\bibitem[{\citenamefont{Ogata et~al.}(1991)\citenamefont{Ogata, Iyetomi, and
  Ichimaru}}]{Ogata1991}
\bibinfo{author}{\bibfnamefont{S.}~\bibnamefont{Ogata}},
  \bibinfo{author}{\bibfnamefont{H.}~\bibnamefont{Iyetomi}}, \bibnamefont{and}
  \bibinfo{author}{\bibfnamefont{S.}~\bibnamefont{Ichimaru}},
  \bibinfo{journal}{The Astrophysical Journal} \textbf{\bibinfo{volume}{372}},
  \bibinfo{pages}{259} (\bibinfo{year}{1991}).

\bibitem[{\citenamefont{Rosenfeld}(1992)}]{Rosenfeld1992}
\bibinfo{author}{\bibfnamefont{Y.}~\bibnamefont{Rosenfeld}},
  \bibinfo{journal}{Phys. Rev. A} \textbf{\bibinfo{volume}{46}},
  \bibinfo{pages}{1059} (\bibinfo{year}{1992}),
  \urlprefix\url{https://link.aps.org/doi/10.1103/PhysRevA.46.1059}.

\bibitem[{\citenamefont{Ichimaru}(1993)}]{Ichimaru1993}
\bibinfo{author}{\bibfnamefont{S.}~\bibnamefont{Ichimaru}},
  \bibinfo{journal}{Reviews of Modern Physics} \textbf{\bibinfo{volume}{65}},
  \bibinfo{pages}{255} (\bibinfo{year}{1993}).

\bibitem[{\citenamefont{Ogata}(1996)}]{Ogata1996}
\bibinfo{author}{\bibfnamefont{S.}~\bibnamefont{Ogata}},
  \bibinfo{journal}{Phys. Rev. E} \textbf{\bibinfo{volume}{53}},
  \bibinfo{pages}{1094} (\bibinfo{year}{1996}),
  \urlprefix\url{https://link.aps.org/doi/10.1103/PhysRevE.53.1094}.

\bibitem[{\citenamefont{Kitamura}(2000)}]{Kitamur2000}
\bibinfo{author}{\bibfnamefont{H.}~\bibnamefont{Kitamura}},
  \bibinfo{journal}{The Astrophysical Journal} \textbf{\bibinfo{volume}{539}},
  \bibinfo{pages}{888} (\bibinfo{year}{2000}),
  \urlprefix\url{http://stacks.iop.org/0004-637X/539/i=2/a=888}.

\bibitem[{\citenamefont{Yakovlev et~al.}(2006)\citenamefont{Yakovlev, Gasques,
  Afanasjev, Beard, and Wiescher}}]{Yakovlev2006}
\bibinfo{author}{\bibfnamefont{D.~G.} \bibnamefont{Yakovlev}},
  \bibinfo{author}{\bibfnamefont{L.~R.} \bibnamefont{Gasques}},
  \bibinfo{author}{\bibfnamefont{A.~V.} \bibnamefont{Afanasjev}},
  \bibinfo{author}{\bibfnamefont{M.}~\bibnamefont{Beard}}, \bibnamefont{and}
  \bibinfo{author}{\bibfnamefont{M.}~\bibnamefont{Wiescher}},
  \bibinfo{journal}{Phys. Rev. C} \textbf{\bibinfo{volume}{74}},
  \bibinfo{pages}{035803} (\bibinfo{year}{2006}),
  \urlprefix\url{https://link.aps.org/doi/10.1103/PhysRevC.74.035803}.

\bibitem[{\citenamefont{Chugunov and DeWitt}(2009)}]{Chugunov2009}
\bibinfo{author}{\bibfnamefont{A.~I.} \bibnamefont{Chugunov}} \bibnamefont{and}
  \bibinfo{author}{\bibfnamefont{H.~E.} \bibnamefont{DeWitt}},
  \bibinfo{journal}{Phys. Rev. C} \textbf{\bibinfo{volume}{80}},
  \bibinfo{pages}{014611} (\bibinfo{year}{2009}),
  \urlprefix\url{https://link.aps.org/doi/10.1103/PhysRevC.80.014611}.

\bibitem[{\citenamefont{{Potekhin, A. Y.} and {Chabrier,
  G.}}(2012)}]{Potekhin2012}
\bibinfo{author}{\bibnamefont{{Potekhin, A. Y.}}} \bibnamefont{and}
  \bibinfo{author}{\bibnamefont{{Chabrier, G.}}}, \bibinfo{journal}{A\&A}
  \textbf{\bibinfo{volume}{538}}, \bibinfo{pages}{A115} (\bibinfo{year}{2012}),
  \urlprefix\url{https://doi.org/10.1051/0004-6361/201117938}.

\bibitem[{\citenamefont{Taguchi et~al.}(1992)\citenamefont{Taguchi, Yasunami,
  and Mima}}]{Taguchi1992}
\bibinfo{author}{\bibfnamefont{T.}~\bibnamefont{Taguchi}},
  \bibinfo{author}{\bibfnamefont{T.}~\bibnamefont{Yasunami}}, \bibnamefont{and}
  \bibinfo{author}{\bibfnamefont{K.}~\bibnamefont{Mima}},
  \bibinfo{journal}{Phys. Rev. A} \textbf{\bibinfo{volume}{45}},
  \bibinfo{pages}{3913} (\bibinfo{year}{1992}),
  \urlprefix\url{https://link.aps.org/doi/10.1103/PhysRevA.45.3913}.

\bibitem[{\citenamefont{Whitley et~al.}(2015)\citenamefont{Whitley, Alley,
  Cabot, Castor, Nilsen, and DeWitt}}]{WHIT15}
\bibinfo{author}{\bibfnamefont{H.~D.} \bibnamefont{Whitley}},
  \bibinfo{author}{\bibfnamefont{W.~E.} \bibnamefont{Alley}},
  \bibinfo{author}{\bibfnamefont{W.~H.} \bibnamefont{Cabot}},
  \bibinfo{author}{\bibfnamefont{J.~I.} \bibnamefont{Castor}},
  \bibinfo{author}{\bibfnamefont{J.}~\bibnamefont{Nilsen}}, \bibnamefont{and}
  \bibinfo{author}{\bibfnamefont{H.~E.} \bibnamefont{DeWitt}},
  \bibinfo{journal}{Contributions to Plasma Physics}
  \textbf{\bibinfo{volume}{55}}, \bibinfo{pages}{413} (\bibinfo{year}{2015}),
  ISSN \bibinfo{issn}{1521-3986},
  \urlprefix\url{http://dx.doi.org/10.1002/ctpp.201400106}.

\bibitem[{\citenamefont{Cl{\'e}rouin
  et~al.}(2016{\natexlab{a}})\citenamefont{Cl{\'e}rouin, Arnault, Ticknor,
  Kress, and Collins}}]{CLER16}
\bibinfo{author}{\bibfnamefont{J.}~\bibnamefont{Cl{\'e}rouin}},
  \bibinfo{author}{\bibfnamefont{P.}~\bibnamefont{Arnault}},
  \bibinfo{author}{\bibfnamefont{C.}~\bibnamefont{Ticknor}},
  \bibinfo{author}{\bibfnamefont{J.~D.} \bibnamefont{Kress}}, \bibnamefont{and}
  \bibinfo{author}{\bibfnamefont{L.~A.} \bibnamefont{Collins}},
  \bibinfo{journal}{Phys. Rev. Lett.} \textbf{\bibinfo{volume}{116}},
  \bibinfo{pages}{115003} (\bibinfo{year}{2016}{\natexlab{a}}),
  \urlprefix\url{http://link.aps.org/doi/10.1103/PhysRevLett.116.115003}.

\bibitem[{\citenamefont{Hansen and McDonald}(2006)}]{HANS06}
\bibinfo{author}{\bibfnamefont{J.-P.} \bibnamefont{Hansen}} \bibnamefont{and}
  \bibinfo{author}{\bibfnamefont{I.~R.} \bibnamefont{McDonald}},
  \emph{\bibinfo{title}{Theory of simple liquids}}
  (\bibinfo{publisher}{Academic Press Cambridge}, \bibinfo{year}{2006}),
  \bibinfo{edition}{3rd} ed.

\bibitem[{\citenamefont{{Rogers}}(1980)}]{ROGE80}
\bibinfo{author}{\bibfnamefont{F.~J.} \bibnamefont{{Rogers}}},
  \bibinfo{journal}{The Journal of Chemical Physics}
  \textbf{\bibinfo{volume}{73}}, \bibinfo{pages}{6272} (\bibinfo{year}{1980}),
  \eprint{https://doi.org/10.1063/1.440124},
  \urlprefix\url{https://doi.org/10.1063/1.440124}.

\bibitem[{\citenamefont{Kravchuk and Yakovlev}(2014)}]{Kravchuk2014}
\bibinfo{author}{\bibfnamefont{P.~A.} \bibnamefont{Kravchuk}} \bibnamefont{and}
  \bibinfo{author}{\bibfnamefont{D.~G.} \bibnamefont{Yakovlev}},
  \bibinfo{journal}{Phys. Rev. C} \textbf{\bibinfo{volume}{89}},
  \bibinfo{pages}{015802} (\bibinfo{year}{2014}),
  \urlprefix\url{https://link.aps.org/doi/10.1103/PhysRevC.89.015802}.

\bibitem[{\citenamefont{Jancovici}(1977)}]{Jancovici1977}
\bibinfo{author}{\bibfnamefont{B.}~\bibnamefont{Jancovici}},
  \bibinfo{journal}{Journal of Statistical Physics}
  \textbf{\bibinfo{volume}{17}}, \bibinfo{pages}{357} (\bibinfo{year}{1977}),
  ISSN \bibinfo{issn}{1572-9613},
  \urlprefix\url{https://doi.org/10.1007/BF01014403}.

\bibitem[{\citenamefont{Widom}(1963)}]{Widom1963}
\bibinfo{author}{\bibfnamefont{B.}~\bibnamefont{Widom}}, \bibinfo{journal}{The
  Journal of Chemical Physics} \textbf{\bibinfo{volume}{39}},
  \bibinfo{pages}{2808} (\bibinfo{year}{1963}),
  \eprint{https://doi.org/10.1063/1.1734110},
  \urlprefix\url{https://doi.org/10.1063/1.1734110}.

\bibitem[{\citenamefont{Ogata}(1997)}]{Ogat1997}
\bibinfo{author}{\bibfnamefont{S.}~\bibnamefont{Ogata}}, \bibinfo{journal}{The
  Astrophysical Journal} \textbf{\bibinfo{volume}{481}}, \bibinfo{pages}{883}
  (\bibinfo{year}{1997}),
  \urlprefix\url{http://stacks.iop.org/0004-637X/481/i=2/a=883}.

\bibitem[{\citenamefont{Pollock and Militzer}(2004)}]{Pollock2004}
\bibinfo{author}{\bibfnamefont{E.~L.} \bibnamefont{Pollock}} \bibnamefont{and}
  \bibinfo{author}{\bibfnamefont{B.}~\bibnamefont{Militzer}},
  \bibinfo{journal}{Phys. Rev. Lett.} \textbf{\bibinfo{volume}{92}},
  \bibinfo{pages}{021101} (\bibinfo{year}{2004}),
  \urlprefix\url{https://link.aps.org/doi/10.1103/PhysRevLett.92.021101}.

\bibitem[{\citenamefont{Militzer and Pollock}(2005)}]{Militzer2005}
\bibinfo{author}{\bibfnamefont{B.}~\bibnamefont{Militzer}} \bibnamefont{and}
  \bibinfo{author}{\bibfnamefont{E.~L.} \bibnamefont{Pollock}},
  \bibinfo{journal}{Phys. Rev. B} \textbf{\bibinfo{volume}{71}},
  \bibinfo{pages}{134303} (\bibinfo{year}{2005}),
  \urlprefix\url{https://link.aps.org/doi/10.1103/PhysRevB.71.134303}.

\bibitem[{\citenamefont{Chugunov et~al.}(2007)\citenamefont{Chugunov, DeWitt,
  and Yakovlev}}]{Chugunov2007}
\bibinfo{author}{\bibfnamefont{A.~I.} \bibnamefont{Chugunov}},
  \bibinfo{author}{\bibfnamefont{H.~E.} \bibnamefont{DeWitt}},
  \bibnamefont{and} \bibinfo{author}{\bibfnamefont{D.~G.}
  \bibnamefont{Yakovlev}}, \bibinfo{journal}{Phys. Rev. D}
  \textbf{\bibinfo{volume}{76}}, \bibinfo{pages}{025028}
  (\bibinfo{year}{2007}),
  \urlprefix\url{https://link.aps.org/doi/10.1103/PhysRevD.76.025028}.

\bibitem[{\citenamefont{Brown and Sawyer}(1997)}]{BROW97}
\bibinfo{author}{\bibfnamefont{L.~S.} \bibnamefont{Brown}} \bibnamefont{and}
  \bibinfo{author}{\bibfnamefont{R.~F.} \bibnamefont{Sawyer}},
  \bibinfo{journal}{Rev. Mod. Phys.} \textbf{\bibinfo{volume}{69}},
  \bibinfo{pages}{411} (\bibinfo{year}{1997}),
  \urlprefix\url{https://link.aps.org/doi/10.1103/RevModPhys.69.411}.

\bibitem[{\citenamefont{More}(1983)}]{MORE83}
\bibinfo{author}{\bibfnamefont{R.~M.} \bibnamefont{More}}, \bibinfo{type}{Tech.
  Rep.} \bibinfo{number}{UCRL-84991}, \bibinfo{institution}{Lawrence Livermore
  Laboratory} (\bibinfo{year}{1983}).

\bibitem[{\citenamefont{More}(1985)}]{MORE85}
\bibinfo{author}{\bibfnamefont{R.~M.} \bibnamefont{More}}, in
  \emph{\bibinfo{booktitle}{Advances in atomic and molecular physics}}
  (\bibinfo{publisher}{Academic Press}, \bibinfo{year}{1985}),
  vol.~\bibinfo{volume}{21}, pp. \bibinfo{pages}{305--355}.

\bibitem[{\citenamefont{Cl{\'e}rouin et~al.}(2013)\citenamefont{Cl{\'e}rouin,
  Robert, Arnault, Kress, and Collins}}]{CLER13b}
\bibinfo{author}{\bibfnamefont{J.}~\bibnamefont{Cl{\'e}rouin}},
  \bibinfo{author}{\bibfnamefont{G.}~\bibnamefont{Robert}},
  \bibinfo{author}{\bibfnamefont{P.}~\bibnamefont{Arnault}},
  \bibinfo{author}{\bibfnamefont{J.~D.} \bibnamefont{Kress}}, \bibnamefont{and}
  \bibinfo{author}{\bibfnamefont{L.~A.} \bibnamefont{Collins}},
  \bibinfo{journal}{Phys. Rev. E} \textbf{\bibinfo{volume}{87}},
  \bibinfo{pages}{061101} (\bibinfo{year}{2013}),
  \urlprefix\url{http://link.aps.org/doi/10.1103/PhysRevE.87.061101}.

\bibitem[{\citenamefont{Arnault et~al.}(2013)\citenamefont{Arnault,
  {Cl{\'e}rouin}, Robert, Ticknor, Kress, and Collins}}]{ARNA13}
\bibinfo{author}{\bibfnamefont{P.}~\bibnamefont{Arnault}},
  \bibinfo{author}{\bibfnamefont{J.}~\bibnamefont{{Cl{\'e}rouin}}},
  \bibinfo{author}{\bibfnamefont{G.}~\bibnamefont{Robert}},
  \bibinfo{author}{\bibfnamefont{C.}~\bibnamefont{Ticknor}},
  \bibinfo{author}{\bibfnamefont{J.~D.} \bibnamefont{Kress}}, \bibnamefont{and}
  \bibinfo{author}{\bibfnamefont{L.~A.} \bibnamefont{Collins}},
  \bibinfo{journal}{Phys. Rev. E} \textbf{\bibinfo{volume}{88}},
  \bibinfo{pages}{063106} (\bibinfo{year}{2013}),
  \urlprefix\url{http://link.aps.org/doi/10.1103/PhysRevE.88.063106}.

\bibitem[{\citenamefont{Cl{\'e}rouin et~al.}(2015)\citenamefont{Cl{\'e}rouin,
  Robert, Arnault, Ticknor, Kress, and Collins}}]{CLER15}
\bibinfo{author}{\bibfnamefont{J.}~\bibnamefont{Cl{\'e}rouin}},
  \bibinfo{author}{\bibfnamefont{G.}~\bibnamefont{Robert}},
  \bibinfo{author}{\bibfnamefont{P.}~\bibnamefont{Arnault}},
  \bibinfo{author}{\bibfnamefont{C.}~\bibnamefont{Ticknor}},
  \bibinfo{author}{\bibfnamefont{J.~D.} \bibnamefont{Kress}}, \bibnamefont{and}
  \bibinfo{author}{\bibfnamefont{L.~A.} \bibnamefont{Collins}},
  \bibinfo{journal}{Phys. Rev. E.} \textbf{\bibinfo{volume}{91}},
  \bibinfo{pages}{011101(R)} (\bibinfo{year}{2015}),
  \urlprefix\url{http://link.aps.org/doi/10.1103/PhysRevE.91.011101}.

\bibitem[{\citenamefont{Cl{\'e}rouin
  et~al.}(2016{\natexlab{b}})\citenamefont{Cl{\'e}rouin, Desbiens, Dubois, and
  Arnault}}]{CLER16b}
\bibinfo{author}{\bibfnamefont{J.}~\bibnamefont{Cl{\'e}rouin}},
  \bibinfo{author}{\bibfnamefont{N.}~\bibnamefont{Desbiens}},
  \bibinfo{author}{\bibfnamefont{V.}~\bibnamefont{Dubois}}, \bibnamefont{and}
  \bibinfo{author}{\bibfnamefont{P.}~\bibnamefont{Arnault}},
  \bibinfo{journal}{Phys. Rev. E} \textbf{\bibinfo{volume}{94}},
  \bibinfo{pages}{061202} (\bibinfo{year}{2016}{\natexlab{b}}),
  \urlprefix\url{http://link.aps.org/doi/10.1103/PhysRevE.94.061202}.

\bibitem[{\citenamefont{{Cl{\'e}rouin}
  et~al.}(1992)\citenamefont{{Cl{\'e}rouin}, {Pollock}, and
  {Z{\'e}rah}}}]{CLER92}
\bibinfo{author}{\bibfnamefont{J.}~\bibnamefont{{Cl{\'e}rouin}}},
  \bibinfo{author}{\bibfnamefont{E.~L.} \bibnamefont{{Pollock}}},
  \bibnamefont{and}
  \bibinfo{author}{\bibfnamefont{G.}~\bibnamefont{{Z{\'e}rah}}},
  \bibinfo{journal}{Phys. Rev. A} \textbf{\bibinfo{volume}{46}},
  \bibinfo{pages}{5130} (\bibinfo{year}{1992}).

\bibitem[{\citenamefont{Lambert et~al.}(2007)\citenamefont{Lambert,
  Cl{\'e}rouin, Mazevet, and Gilles}}]{LAMB07}
\bibinfo{author}{\bibfnamefont{F.}~\bibnamefont{Lambert}},
  \bibinfo{author}{\bibfnamefont{J.}~\bibnamefont{Cl{\'e}rouin}},
  \bibinfo{author}{\bibfnamefont{S.}~\bibnamefont{Mazevet}}, \bibnamefont{and}
  \bibinfo{author}{\bibfnamefont{D.}~\bibnamefont{Gilles}},
  \bibinfo{journal}{Contributions to Plasma Physics}
  \textbf{\bibinfo{volume}{47}}, \bibinfo{pages}{272} (\bibinfo{year}{2007}),
  \bibinfo{note}{dOI: 10.1002/ctpp.200710037},
  \urlprefix\url{http://adsabs.harvard.edu/abs/2007CoPP...47..272L}.

\bibitem[{\citenamefont{Perrot}(1979)}]{PERR79}
\bibinfo{author}{\bibfnamefont{F.}~\bibnamefont{Perrot}},
  \bibinfo{journal}{Phys. Rev. A} \textbf{\bibinfo{volume}{20}},
  \bibinfo{pages}{586} (\bibinfo{year}{1979}).

\bibitem[{\citenamefont{Ticknor et~al.}(2016)\citenamefont{Ticknor, Kress,
  Collins, Cl{\'e}rouin, Arnault, and Decoster}}]{TICK16}
\bibinfo{author}{\bibfnamefont{C.}~\bibnamefont{Ticknor}},
  \bibinfo{author}{\bibfnamefont{J.~D.} \bibnamefont{Kress}},
  \bibinfo{author}{\bibfnamefont{L.~A.} \bibnamefont{Collins}},
  \bibinfo{author}{\bibfnamefont{J.}~\bibnamefont{Cl{\'e}rouin}},
  \bibinfo{author}{\bibfnamefont{P.}~\bibnamefont{Arnault}}, \bibnamefont{and}
  \bibinfo{author}{\bibfnamefont{A.}~\bibnamefont{Decoster}},
  \bibinfo{journal}{Phys. Rev. E} \textbf{\bibinfo{volume}{93}},
  \bibinfo{pages}{063208} (\bibinfo{year}{2016}),
  \urlprefix\url{http://link.aps.org/doi/10.1103/PhysRevE.93.063208}.

\bibitem[{\citenamefont{Ichimaru et~al.}(1984)\citenamefont{Ichimaru, Tanaka,
  and Iyetomi}}]{ICHI84}
\bibinfo{author}{\bibfnamefont{S.}~\bibnamefont{Ichimaru}},
  \bibinfo{author}{\bibfnamefont{S.}~\bibnamefont{Tanaka}}, \bibnamefont{and}
  \bibinfo{author}{\bibfnamefont{H.}~\bibnamefont{Iyetomi}},
  \bibinfo{journal}{Phys. Rev. A} \textbf{\bibinfo{volume}{29}},
  \bibinfo{pages}{2033} (\bibinfo{year}{1984}).

\bibitem[{\citenamefont{Cl{\'e}rouin et~al.}(2017)\citenamefont{Cl{\'e}rouin,
  Arnault, Desbiens, White, Ticknor, Kress, and Collins}}]{CLER17}
\bibinfo{author}{\bibfnamefont{J.}~\bibnamefont{Cl{\'e}rouin}},
  \bibinfo{author}{\bibfnamefont{P.}~\bibnamefont{Arnault}},
  \bibinfo{author}{\bibfnamefont{N.}~\bibnamefont{Desbiens}},
  \bibinfo{author}{\bibfnamefont{A.~J.} \bibnamefont{White}},
  \bibinfo{author}{\bibfnamefont{C.}~\bibnamefont{Ticknor}},
  \bibinfo{author}{\bibfnamefont{J.~D.} \bibnamefont{Kress}}, \bibnamefont{and}
  \bibinfo{author}{\bibfnamefont{L.~A.} \bibnamefont{Collins}},
  \bibinfo{journal}{Contrib. Plasma Phys.} \textbf{\bibinfo{volume}{57}},
  \bibinfo{pages}{512} (\bibinfo{year}{2017}).

\bibitem[{\citenamefont{Arnault}(2013)}]{ARNA13b}
\bibinfo{author}{\bibfnamefont{P.}~\bibnamefont{Arnault}},
  \bibinfo{journal}{High Energy Density Physics} \textbf{\bibinfo{volume}{9}},
  \bibinfo{pages}{711} (\bibinfo{year}{2013}),
  \urlprefix\url{http://www.sciencedirect.com/science/article/pii/S1574181813001651}.

\bibitem[{\citenamefont{White et~al.}(2017)\citenamefont{White, Collins, Kress,
  Ticknor, Cl{\'e}rouin, Arnault, and Desbiens}}]{WHIT17}
\bibinfo{author}{\bibfnamefont{A.~J.} \bibnamefont{White}},
  \bibinfo{author}{\bibfnamefont{L.~A.} \bibnamefont{Collins}},
  \bibinfo{author}{\bibfnamefont{J.~D.} \bibnamefont{Kress}},
  \bibinfo{author}{\bibfnamefont{C.}~\bibnamefont{Ticknor}},
  \bibinfo{author}{\bibfnamefont{J.}~\bibnamefont{Cl{\'e}rouin}},
  \bibinfo{author}{\bibfnamefont{P.}~\bibnamefont{Arnault}}, \bibnamefont{and}
  \bibinfo{author}{\bibfnamefont{N.}~\bibnamefont{Desbiens}},
  \bibinfo{journal}{Phys. Rev. E} \textbf{\bibinfo{volume}{95}},
  \bibinfo{pages}{063202} (\bibinfo{year}{2017}),
  \urlprefix\url{https://link.aps.org/doi/10.1103/PhysRevE.95.063202}.

\end{thebibliography}

\clearpage

\appendix*
\renewcommand\thefigure{A.\arabic{figure}}

\section{Structures and correlations}
\setcounter{figure}{0}


%
\begin{sidewaysfigure}
\vspace{6cm}
\includegraphics[width=25 cm]{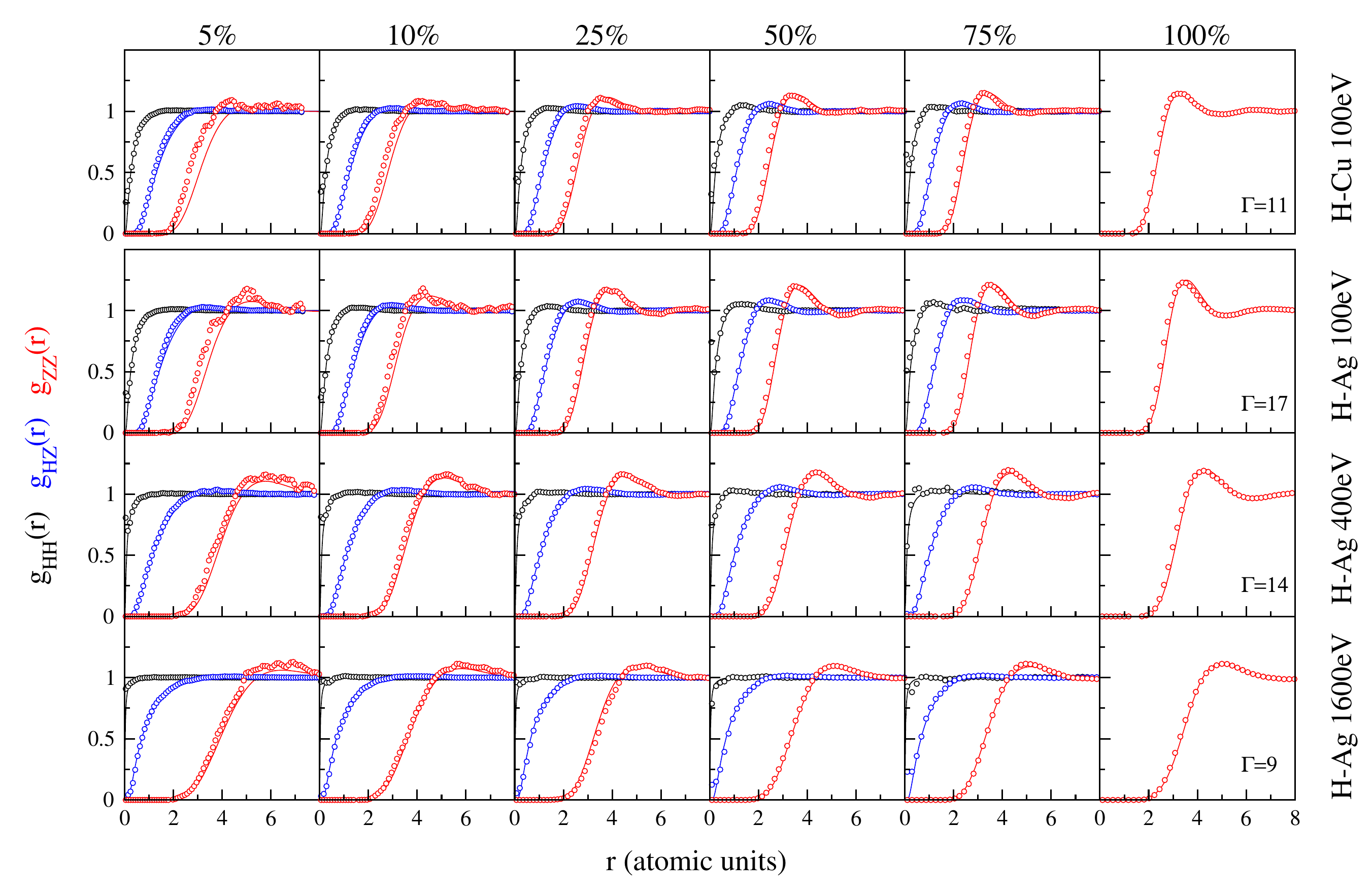}
\caption{\label{multi} Overview of the pair distribution functions ($g_{HH}$: black, $g_{HZ}$: blue and $g_{ZZ}$: red) obtained by  OFMD simulations (circles) compared to corresponding MCHNC calculations (lines). The heavy component (Cu or Ag) concentration increases from 5\% to 100\% left to right. First row: H-Cu mixtures at 100\,eV. Second, third and fourth rows: H-Ag mixtures at respectively 100, 400 and 1600\,eV  (see Table\,\ref{tab1}).}
\end{sidewaysfigure}

\end{document}